\documentclass[11pt]{article}
\usepackage{cospar-umd}
\usepackage{url}

\usepackage{graphicx}
\usepackage[figuresright]{rotating}

\title{ MIGRATION OF INTERPLANETARY DUST}

\author{S. I. Ipatov\address{(1) 
George Mason University, USA;
(2) NASA/GSFC, Mail Code 685, 
Greenbelt, MD 20771, USA; 
E-mail: siipatov@hotmail.com  (for correspondence); 
(3) Institute of Applied Mathematics, 
Moscow, Russia}, 
J. C. Mather\address{NASA/GSFC, Mail Code 685, Greenbelt,
MD 20771, USA},
P. A. Taylor
\address{University of Maryland, College Park, MD 20742, USA}
}

\begin{document}

\maketitle

\begin{abstract}

We numerically investigate the migration of dust particles with initial orbits close to those of 
the numbered asteroids, observed trans-Neptunian objects, and Comet Encke. The fraction
of silicate asteroidal particles that collided 
with the Earth during their lifetime varied from 1.1\% for 100 micron particles 
to 0.008\% for 1 micron particles. Almost all asteroidal particles with diameter $d$$\ge$4 
microns collided with the Sun. 
The peaks in the migrating asteroidal dust particles' semi-major axis distribution
at the n:(n+1) resonances with Earth and Venus 
and the gaps associated with
the 1:1 resonances with these planets
are more pronounced for larger particles.
The probability of collisions of cometary particles with the Earth
is smaller than for asteroidal particles, and this difference is greater
for larger particles.

\end{abstract}

\section*{INTRODUCTION}

The orbital evolution of interplanetary dust particles originating from the trans-Neptunian belt 
[1]-[5], Halley- and Encke-type comets [6],[7], and asteroid families and short-period comets 
[8]-[10] has been numerically simulated by several authors in order to understand the 
distribution of dust in the solar system.  Observationally, the infrared satellites IRAS and 
COBE detected thermal emission from dust outside the Earth's orbit [11]-[13] while Pioneers 
10 and 11 measured the interplanetary dust population directly using impact detectors [14].
The main contribution to the zodiacal light is from particles with diameters of about 20 to 
200 $\mu$m. It has been estimated that 20,000 to 40,000 tons of micrometer- to 
millimeter-sized dust particles fall to the Earth every year with the mass distribution 
of dust particles 
peaking at about 200 $\mu$m in diameter [15].
Liou et al. [7] showed that the observed shape of the zodiacal cloud 
can be accounted for by a combination of about 1/4 to 1/3 
asteroidal dust and about 3/4 to 2/3 cometary dust.
The hypothesis that the 'kuiperoidal' dust (i.e., the dust formed in the 
Edgeworth-Kuiper belt) is dominant in the outer solar system explains the Pioneer and 
Voyager data fairly well, while fitting a sample of the COBE/DIRBE data indicates that 
the kuiperoidal dust contributes as much as 1/3 of the total number density near the 
Earth (for a review, see [4]). Ozernoy [4] considered 1 m and 5 m 
particles while constructing the brightness of a disk of asteroidal, cometary, and 
kuiperoidal dust particles which fit the COBE/DIRBE data. 


Analysis of the Pioneer 10 and 11 meteoroid detector data [16], [17] showed
that a population of $10^{-9}$ and $10^{-8}$ g ($\sim$10 $\mu$m)
particles has a constant spatial density
between 3 and 18 AU. Dust grains released by Halley-type comets cannot account for this 
observed distribution [6], but trans-Neptunian dust particles can [2].
Using trans-Neptunian dust particles, Moro-Martin and Malhotra [5, Fig. 13] found that the 
surface (not spatial) density was approximately constant in this region. 

Liou et al. [1] noted that interstellar dust particles with an average size of 1.2 $\mu$m
can destroy dust particles formed in the solar system
and that the collisional lifetimes for 1, 2, 4, 9, 23 $\mu$m particles 
are 104, 49, 19, 4.8, 0.86 Myr, respectively.
In these size ranges mutual collisions are not as important as collisions
with interstellar grains.
Moro-Martin and Malhotra [5] concluded that collisional destruction
is most important for kuiperoidal grains between 6 $\mu$m 
(9 $\mu$m in Liou et al.) and 50 $\mu$m. Particles larger than 50 $\mu$m may survive 
because interstellar grains are too
small to destroy them in a single impact. For silicate
particles 1-40 $\mu$m in diameter, the sublimation temperature ($\sim$1500 K) is
reached at $R$$<$0.5 AU, but for water ice particles the sublimation temperature 
($\sim$100 K) is reached at 27, 19, 14, 10, and 4.3 AU for the sizes of 3, 6, 11, 23, 
and 120 $\mu$m, respectively [1],[5].

In the present paper we consider a different set of initial orbits for the dust particles 
than the aforementioned authors
and, for the first time, investigate the collisional probabilities of migrating particles 
with the planets based on a set of orbital elements during their evolution. We also present
plots of the orbital elements of the migrating particles.

\section*{MODELS FOR ASTEROIDAL DUST PARTICLES}

    Using the Bulirsh--Stoer method of integration, we investigated
the migration of asteroidal dust particles under the influence of planetary gravity
(excluding Pluto), radiation pressure, Poynting--Robertson drag, and solar wind 
drag for values of
the ratio between the radiation pressure force and the gravitational force
$\beta$ equal to 0.004, 0.01, 0.05, 0.1, 0.25, and 0.4. 
Burns et al. [18] obtained $\beta$=$0.573 Q_{pr}/(\rho s)$, where
$\rho$ is the particle's density in grams per cubic centimeter,
$s$ is its radius in micrometers, and $Q_{pr}$ is the radiation
pressure coefficient ($Q_{pr}$ is close to unity for particles
larger than 1 $\mu$m). For silicates,
the $\beta$ values 0.004, 0.01, 0.05, 0.1, 0.25, and 0.4 correspond to particle diameters of about 
100, 40, 9, 4, 1.6, and 1 microns, respectively. Silicate particles with $\beta$
values of 0.01 and 0.05 have masses of $10^{-7}$ g and $10^{-9}$ g. 
For water ice, our $\beta$ values correspond to particle
diameters of 300, 120, 23, 11, 6, and 3 $\mu$m [18].
We assume the ratio of solar wind drag 
to Poynting--Robertson drag to be 0.35 [5], [6].
The relative error per integration step was taken to be less than $10^{-8}$.

The initial positions and velocities of the asteroidal particles 
were the same as those of the first numbered main-belt asteroids (JDT 2452500.5),
i.e., dust particles are assumed to leave the asteroids with zero
relative velocity. 
We considered $N$=500 particles for each $\beta$$\ge$0.05,  $N$=250 for $\beta$=0.01,
and $N$=100 for $\beta$=0.004. With $\beta$$\ge$0.01 in each run we took $N$=250, because for $N$$\ge$500 
the computer time per calculation for one particle was several times greater 
than for $N$=250. The simulations continued until all of the particles
either collided with the Sun or reached 2000 AU from the Sun. 
The lifetimes of all considered asteroidal dust particles were less than 0.8 Myr, except for
one particle with a lifetime of 19 Myr.  
We also made similar runs without planets to investigate the role of planets in 
interplanetary dust migration.

\section*{COLLISIONS of ASTEROIDAL  PARTICLES WITH PLANETS AND THE SUN}

In our runs, planets were 
considered as material points, but using orbital elements 
obtained with a step $d_t$ of $\le$20 yr ($d_t$=10 yr for $\beta$ equal to 0.1 
and 0.25, and $d_t$=20 yr for other values of $\beta$), similar to [19] 
we calculated the mean probability 
$P$=$P_{\Sigma}/N$ ($P_\Sigma$ is the probability for all $N$ 
considered particles) of a collision of a particle with a planet 
during the lifetime of the particle. We define
$T$=$T_{\Sigma}/N$ as the mean time during which the perihelion 
distance $q$ of a particle was less than the semi-major axis of the 
planet and
$T_J$ as the mean time spent in Jupiter-crossing orbits.
Below, $P_{Sun}$ is the ratio of the number of particles that
collided with the Sun to the total number of particles.
$T_S^{min}$ and $T_S^{max}$ are the minimum and maximum values of the time until
the  collision of a particle with the Sun, and $T_{2000}^{min}$
and $T_{2000}^{max}$ are the minimum and maximum values of the time 
 when the distance between a particle and the Sun reached 2000 AU.
With $\beta$$\ge$0.01 the values of 
$P_r$=$10^6 P$, $T$, $T_J$,
$T_S^{min}$, $T_S^{max}$, $T_{2000}^{min}$,
and $T_{2000}^{max}$  are shown in 
Table 1 for $N$=250 (for $\beta$=0.1 we present two runs with 250
different particles), and $P_{Sun}$ was obtained for all considered 
particles at a fixed $\beta$.

The minimum time $T_S^{min}$ needed to reach the Sun is smaller for smaller particles, but 
the ratio  $T_S^{max}/T_S^{min}$ is much
greater 
and the ratio $T_{2000}^{max}/T_S^{max}$ is smaller 
for $\beta$$\ge$0.25 than for $\beta$$\le$0.1. 
For $\beta$=0.05, 498 particles collided with the Sun in less than 0.089 Myr, 
but two particles (with initial orbits close to those of the asteroids
361 and 499), which reached 2000 AU, lived for 0.21 Myr and 19.06 Myr, 
respectively. The latter object's perihelion was near Saturn's orbit for a long time. 
At $\beta$$=$0.05  the first 250 particles did not migrate outside Jupiter's orbit,
so $T_J$=0 in Table 1.

\begin{table}
\caption{Values of $T$, $T_J$, $T_S^{min}$, $T_S^{max}$, $T_{2000}^{min}$,
$T_{2000}^{max}$ (in Kyr), 
$P_r$, 
and $P_{Sun}$ 
obtained for asteroidal dust particles for several values of $\beta$
(Venus=V, Earth=E, Mars=M)
}

$ \begin{array}{lcccccccccccc} 

\hline	

  & & $V$ & $V$ & $E$ & $E$ & $M$ & $M$ & && && \\


\beta & P_{Sun}& P_r & T & P_r & T &  P_r & T & T_J & T_S^{min}&T_S^{max}&
T_{2000}^{min}&T_{2000}^{max} \\

\hline
0.004& 1.000& 12783 & 40.5 & 11350 & 90. & 1204 & 220. & 0 &348 &932&& \\
0.01 & 1.000&1534 & 19.2 & 1746 & 44.2 & 127 & 99.9 & 0 &142& 422&& \\
0.05 & 0.996&195  & 4.0  & 190  & 8.1 & 36.7 & 20.5 & 0 &30&  89& &\\
0.1  & 0.990& 141  & 2.4  & 132  & 4.8 & 16.4 & 12.0 & 2.21 &16&44&138& 793 \\
0.1* & 0.990& 366  & 2.4  & 279  & 4.8 & 20.9 & 12.0 & 0.92 &7.2&43&9& 534 \\
0.25 & 0.618&79.2 & 1.4  & 63.8 & 2.9 & 5.60 & 5.9  & 31.7 & 5.9&385&1.6&567 \\
0.4  & 0.316&12.4 & 1.5  & 8.0  & 2.5 & 0.72 & 8.8  & 32.3 &4.3&172&1.7& 288 \\

\hline
\end{array} $ 
\end{table}

For smaller particles (i.e., those with larger $\beta$), $P_{Sun}$ is smaller
and the probability
of collisions of particles with the 
terrestrial planets is smaller. The probability of a collision of a migrating dust particle 
with the Earth for $\beta$=0.004 is greater by a factor of 1420 than for  $\beta$=0.4. 
These probabilities of collisions are in accordance with cratering records in 
lunar material and on the panels of the Long Duration Exposure Facility, 
which showed that the mass distribution of dust particles encountering Earth peaks at 
$d$=200 $\mu$m [9].

\section*{ORBITAL EVOLUTION OF ASTEROIDAL PARTICLES}

Several plots of the distribution of 
migrating asteroidal particles in their orbital elements and  the distribution 
of particles with their distance $R$ from the Sun and their height $h$ above the initial 
plane of the Earth's orbit are presented in Figs. 1-7 (Fig. 8 present the results for
kuiperoidal dust particles).  For Fig. 1-2
the number of  bins in the semi-major axis $a$ is 1000. For the other figures
the number of bins in $a$ or $R$ is 100, and the number
of bins in $e$ or $h$ is usually slightly less than 100. 
The width of a bin in $i$ is 1$^\circ$ (0.5$^\circ$ for Fig. 8). 

In Figs. 1-2, 4-8 to calculate the 
orbital elements we 
included the coefficient (1-$\beta$) on the mass of the Sun,
because of radiation pressure. 
Fig. 3 is the only exception.  Although similar to Fig. 4, Fig. 3
(see also figures in [20])
shows the osculating orbital elements that are calculated using the normal mass of the Sun 
(i.e., to transfer from rectangular coordinates to
orbital elements we used the same formulas as those for massive bodies).
For transformations between semi-major axes and eccentricities
in the above two systems of orbital elements we used the
formulas obtained by Kortenkamp and Dermott [9].
In total, Figs. 3 and 4 are similar, but they differ in some details.
For example, for $\beta$=0.4 in Fig. 4 there are some pairs of $a$ and $e$
corresponding to a perihelion near Saturn's orbit. There are no
such pairs in Fig. 3 for $\beta$=0.4.

\begin{figure}
\includegraphics[width=80mm]{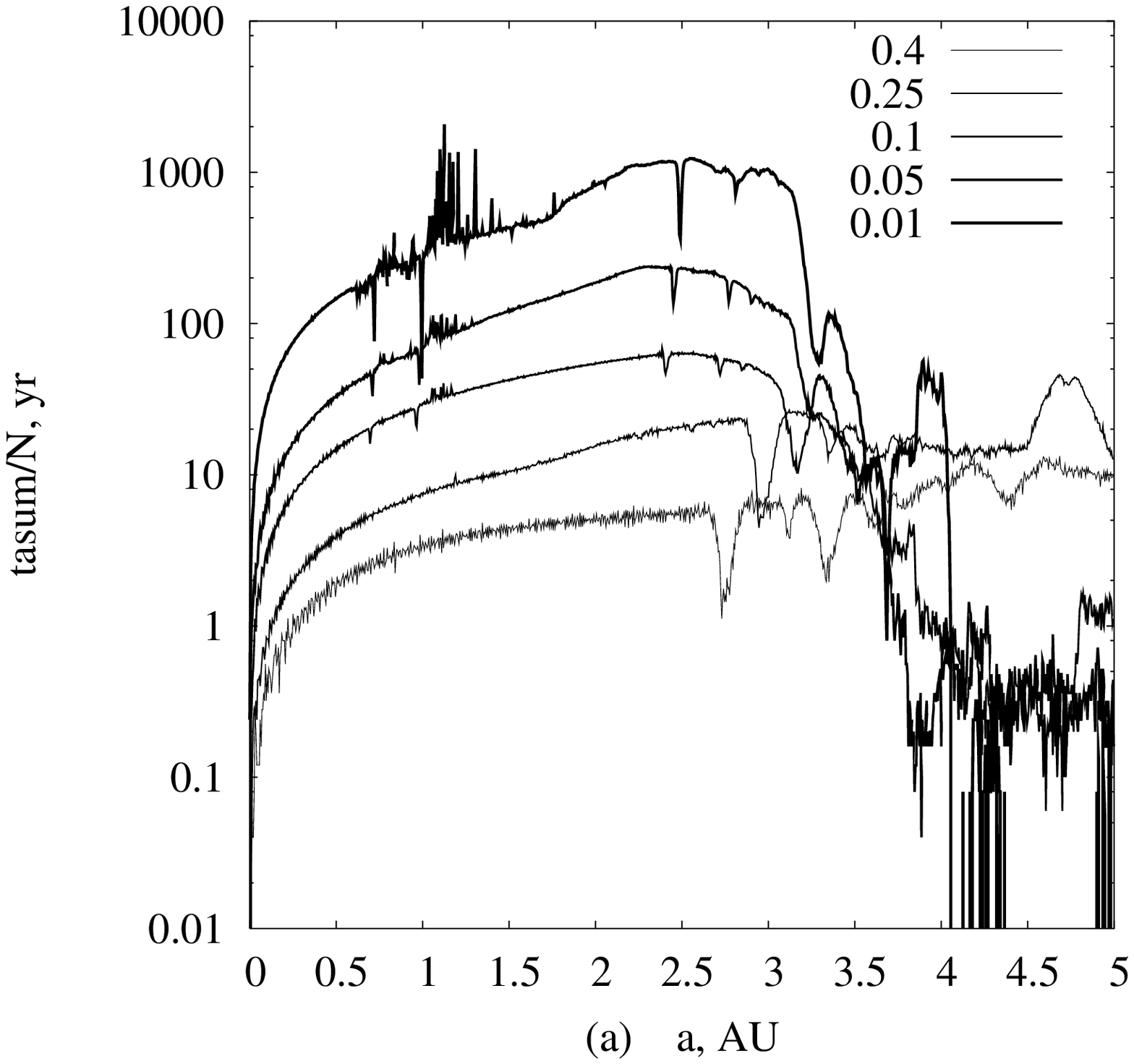}
\includegraphics[width=80mm]{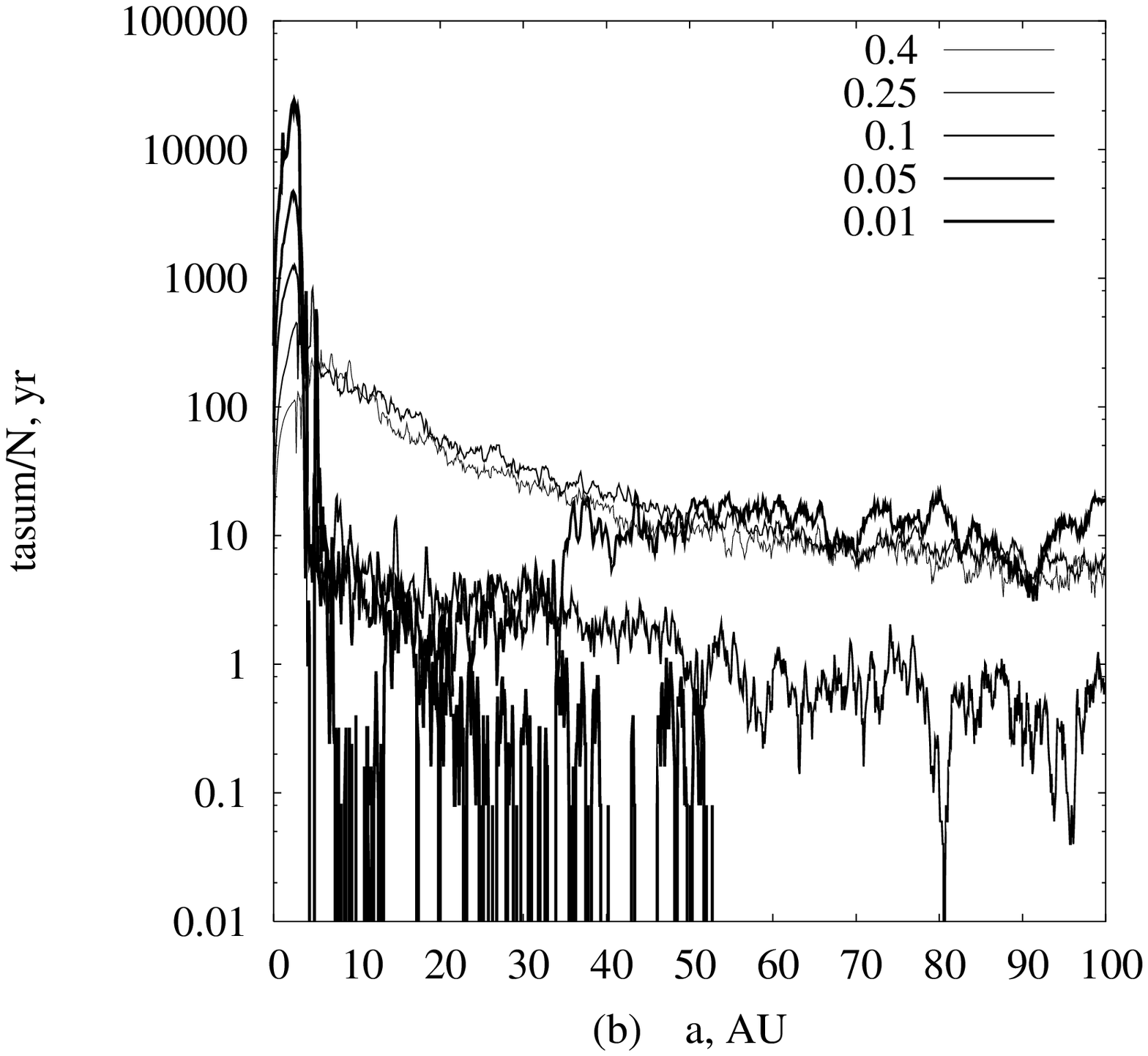}
\includegraphics[width=160mm]{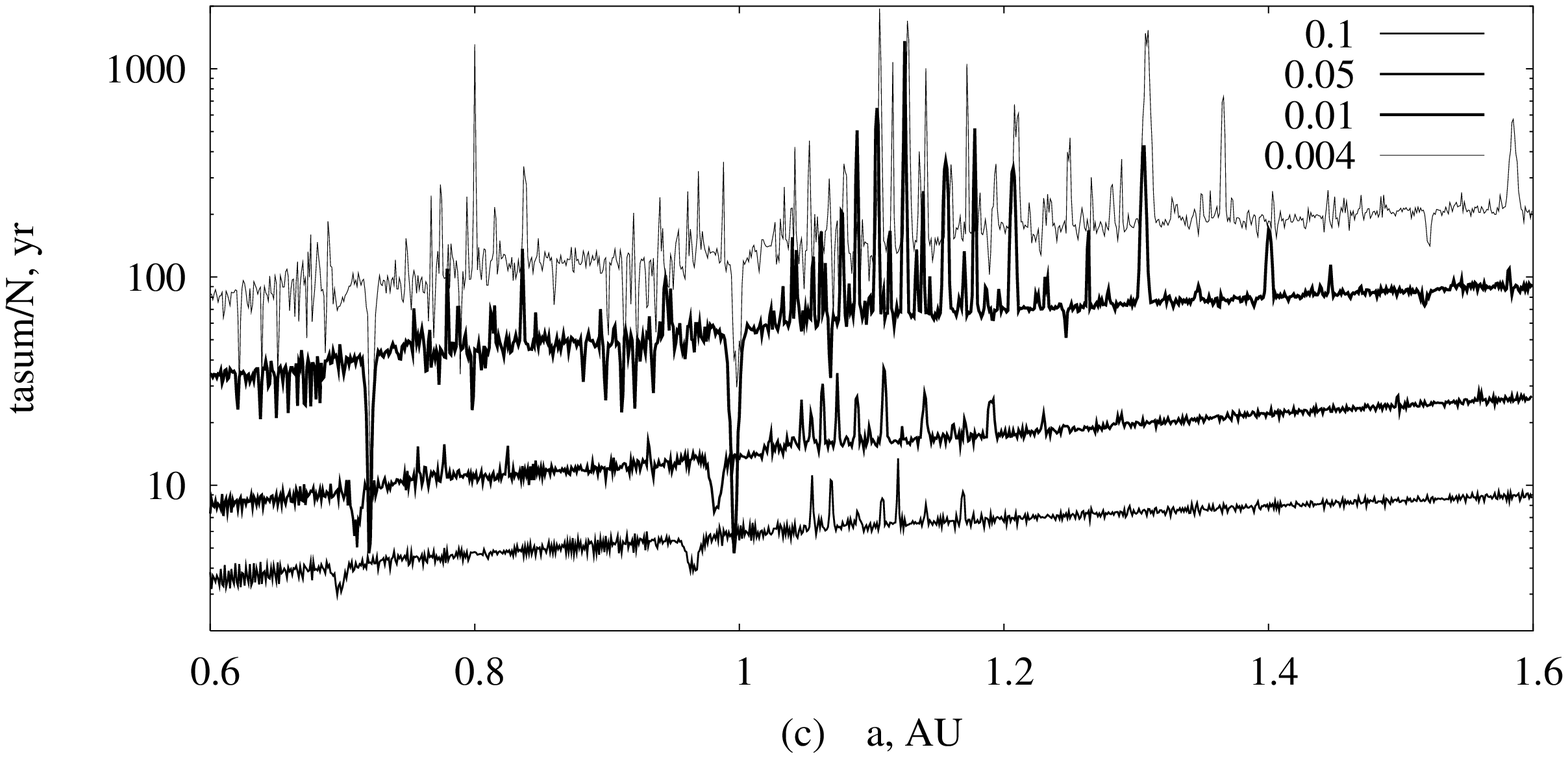}

\caption{
Mean time $t_a$ (the total time {\it tasum} divided by the number 
$N$ of particles) during which an asteroidal dust particle 
had a semi-major axis in an interval with a width of (a) 0.005 AU, (b) 0.1 AU, or (c) 0.001 AU. 
The values of $t_a$ at 1.5 AU are greater for 
smaller $\beta$. Curves plotted at 40 AU are (top-to-bottom) 
for $\beta$ equal to 0.25, 0.4, 0.05, 0.1, and 0.01. For 
calculations of orbital elements we added the coefficient 
(1-$\beta$) to the mass of the Sun, which is due to the radiation 
pressure (the same orbital elements are used
in Figs. 1-2, 4-8). Initial velocities and coordinates of dust particles 
were the same as those of the first $N$=500 numbered asteroids 
($N$=250 for $\beta$=0.01, $N$=100 for $\beta$=0.004) at JDT 2452500.5.
}

\end{figure}%

\begin{figure}
\includegraphics[width=80mm]{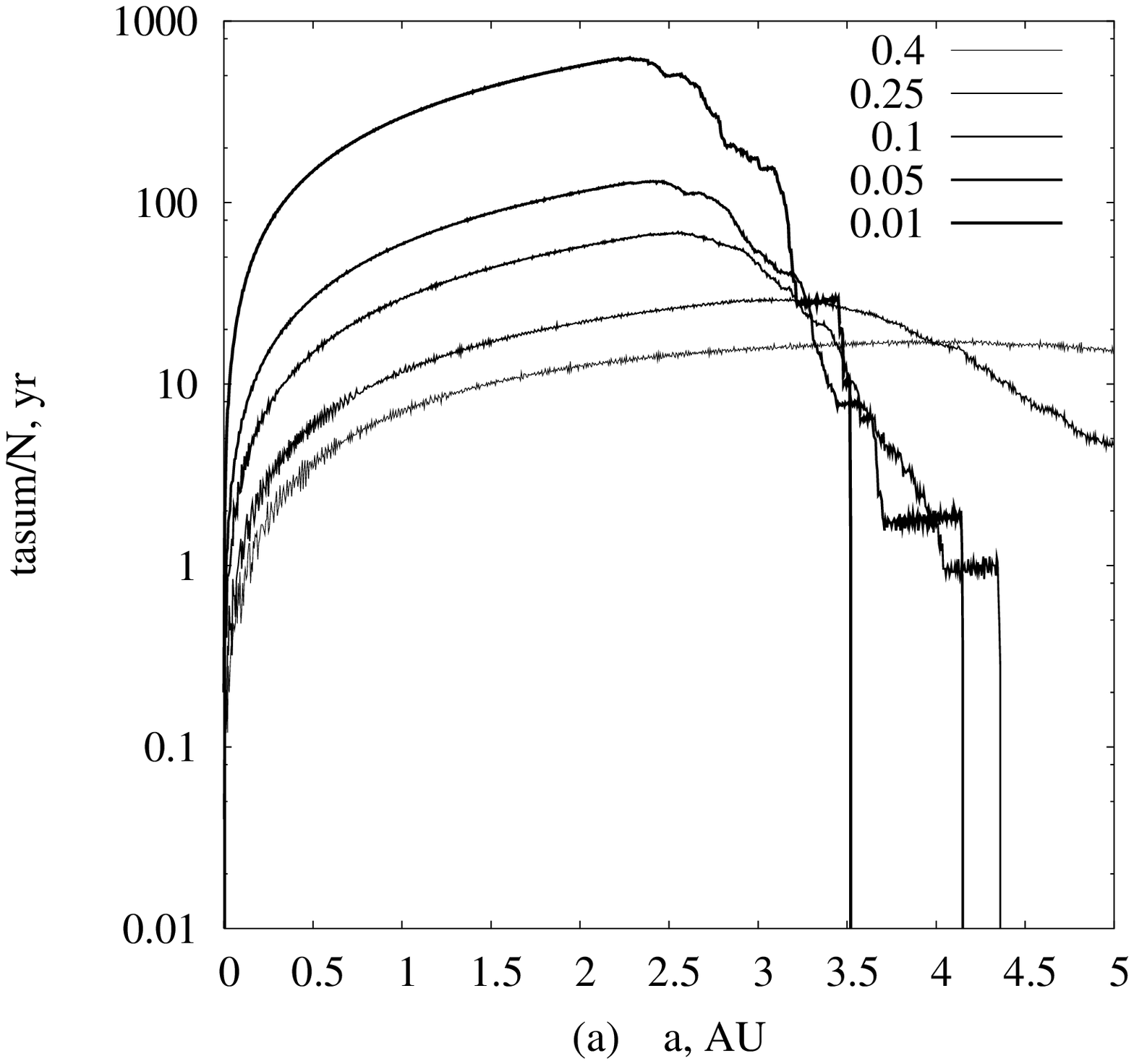}
\includegraphics[width=80mm]{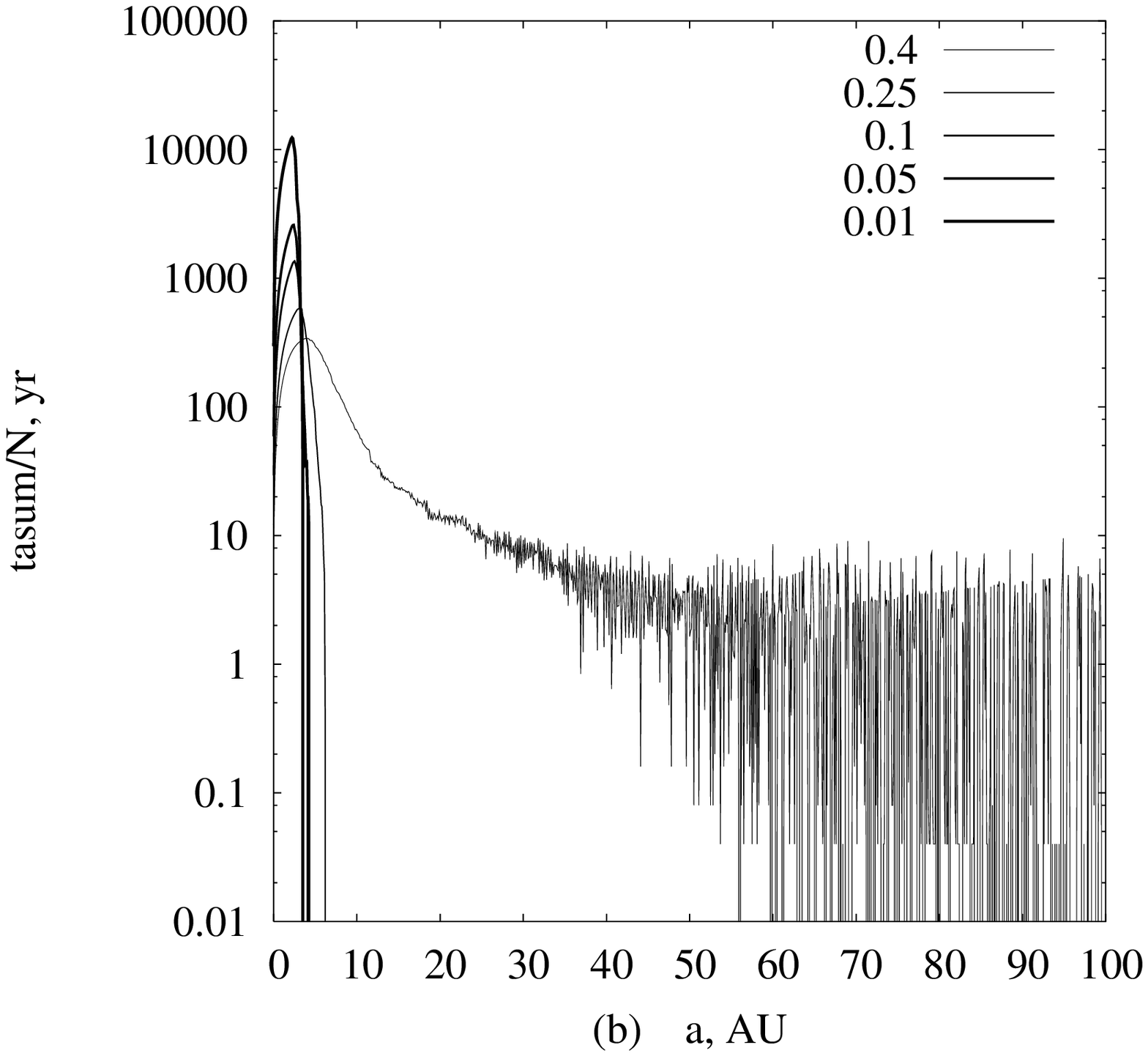}

\caption{
Same as for Fig. 1 but without gravitational influence of 
planets and for a smaller number of initial dust particles 
($N$=250 at $\beta$$\ge$0.05, and $N$=100 at $\beta$=0.01). 
The values of $a$$>$6.2 AU were reached only at $\beta$=0.4. 
}

\end{figure}%

We obtained that the mean time $t_a$ 
(the total time divided by the number $N$ of particles) 
during which an asteroidal dust particle had 
a semi-major axis $a$ in an interval of fixed width is 
greater for smaller $\beta$ at semi-major axes $a$$<$3 AU (exclusive of the gap
at $a$=1 AU and $\beta$=0.01).
In Fig. 1b curves plotted at 40 AU are (top-to-bottom) 
for $\beta$ equal to 0.25, 0.4, 0.05, 0.1, and 0.01.
For $\beta$$\le$0.1 the values 
of $t_a$ are much smaller at $a$$>$3.5 AU than at $1$$<$$a$$<$3 AU, and they are usually a 
maximum at $a$$\approx$2.3 AU. For $\beta$=0.01 the local maxima of $t_a$ 
corresponding to the  6:7, 5:6, 3:4, and 2:3 resonances with the Earth
are greater than the maximum at 2.4 AU. 
There are several other local maxima (Fig. 1c) corresponding to the 
n:(n+1) resonances 
with Earth and Venus (e.g., the 7:8 and 4:5 resonances with Venus).
The trapping of dust particles in the n:(n+1) resonances cause Earth's asteroidal ring 
[21], [22].
The greater the $\beta$, the smaller the local maxima
corresponding to these resonances.
At $\beta$$\le$0.1
there are gaps with $a$ a little smaller than the semi-major 
axes of Venus and Earth that correspond to the 1:1 resonance for each; the greater the 
$\beta$, the smaller the corresponding
values of $a$. 
A small gap for Mars is seen only at $\beta$=0.01.
There are also gaps corresponding to the 3:1, 5:2, and 2:1 resonances
with Jupiter.

For all considered $\beta$, $t_a$ 
decreases considerably  with a decrease of $a$ at $a$$<$1 AU and usually decreases with 
an increase of $a$ at $a$$>$5 AU (Fig. 1). 
Relatively large values of $t_a$ at $a$$>$40 AU for $\beta$=0.05
are due to one particle.
For $a$$>$5 AU the values of $t_a$ are usually a little greater
at $\beta$=0.25 than those at $\beta$=0.4. The number
of particles with $a$$>$5 AU at $\beta$=0.25 is smaller
than at $\beta$=0.4, but they move more slowly to 2000 AU than at 
$\beta$=0.4.
Analyzing Fig. 1a, we can conclude that larger particles make up a greater proportion of the 
dust population in the terrestrial zone than they do in the asteroid belt.

In Fig. 2 we present the results obtained for the model without planets. In this case,
migration outside 5 AU is smaller than in the model with planets 
and, of course, there are no peaks
and gaps caused by planets (see Fig. 1). 
For the model without planets, the values of $T_S^{min}$
were about the same as those presented in Table 1,
but the values of $T_S^{max}$ sometimes were smaller.
With $N$=250 
the values of $P_{Sun}$ for $\beta$=0.25 and $\beta$=0.4 of 0.908 and 0.548,
respectively, are greater than for the model with planets.
For  $\beta$$\le$0.1 all of the particles collided with the Sun.

We now return to the model with planets.
At $a$$<$4 AU the maximum eccentricities for $\beta$$\ge$0.25
were greater than those for $\beta$$\le$0.1 (Figs. 4-5). 
At $\beta$=0.01 some particles migrated into the 1:1 resonance  with Jupiter.
For $a$$>$10 AU perihelia were usually near Jupiter's orbit (for $\beta$=0.05 and $\beta$=0.25 
also near Saturn's orbit). 
In almost all cases, the inclinations $i$$<$$50^o$; at $a$$>$10 AU the
maximum $i$ was smaller for smaller $\beta$ (Fig. 6). 

\begin{figure}
\includegraphics[width=80mm]{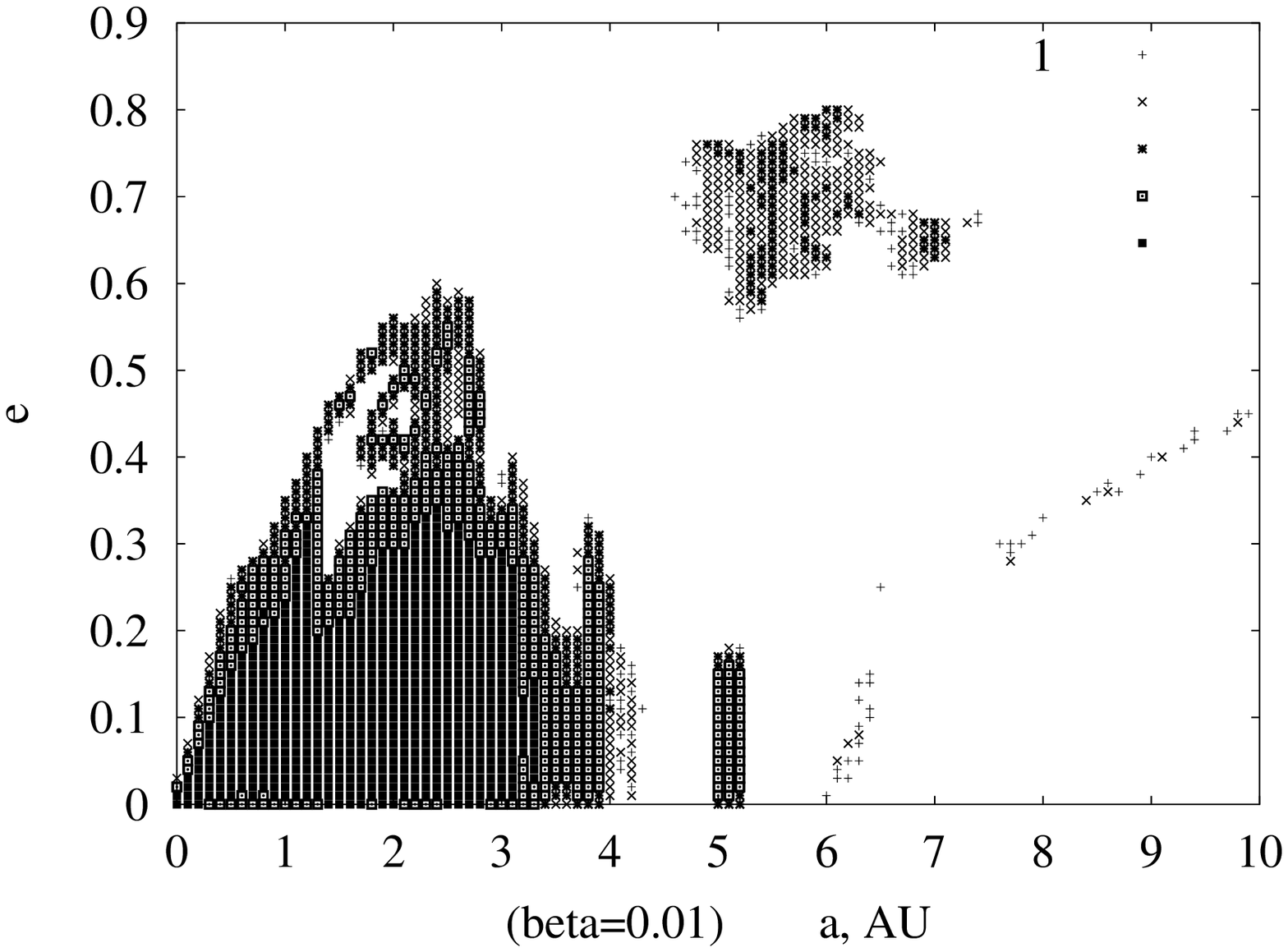}
\includegraphics[width=80mm]{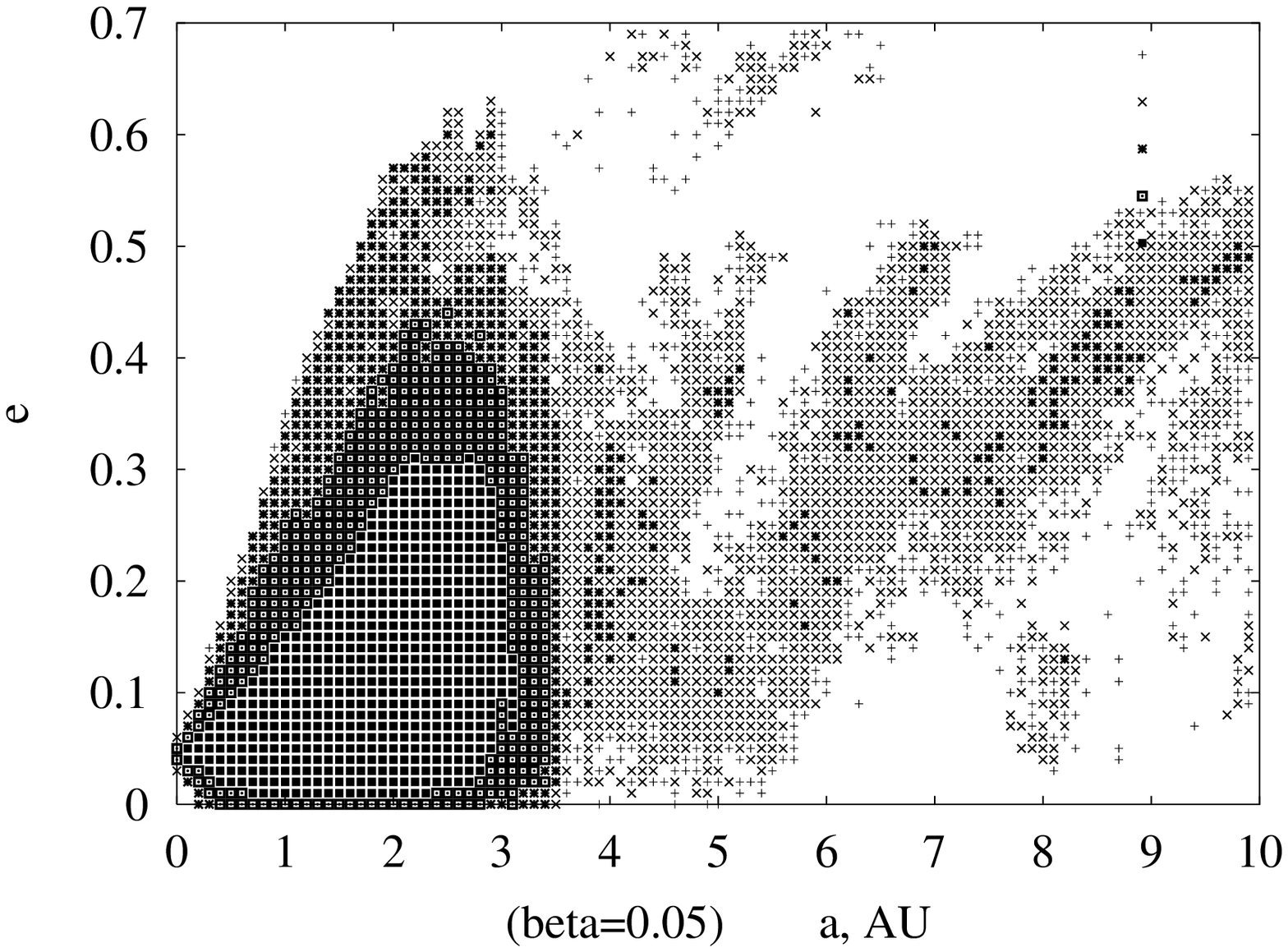}
\includegraphics[width=80mm]{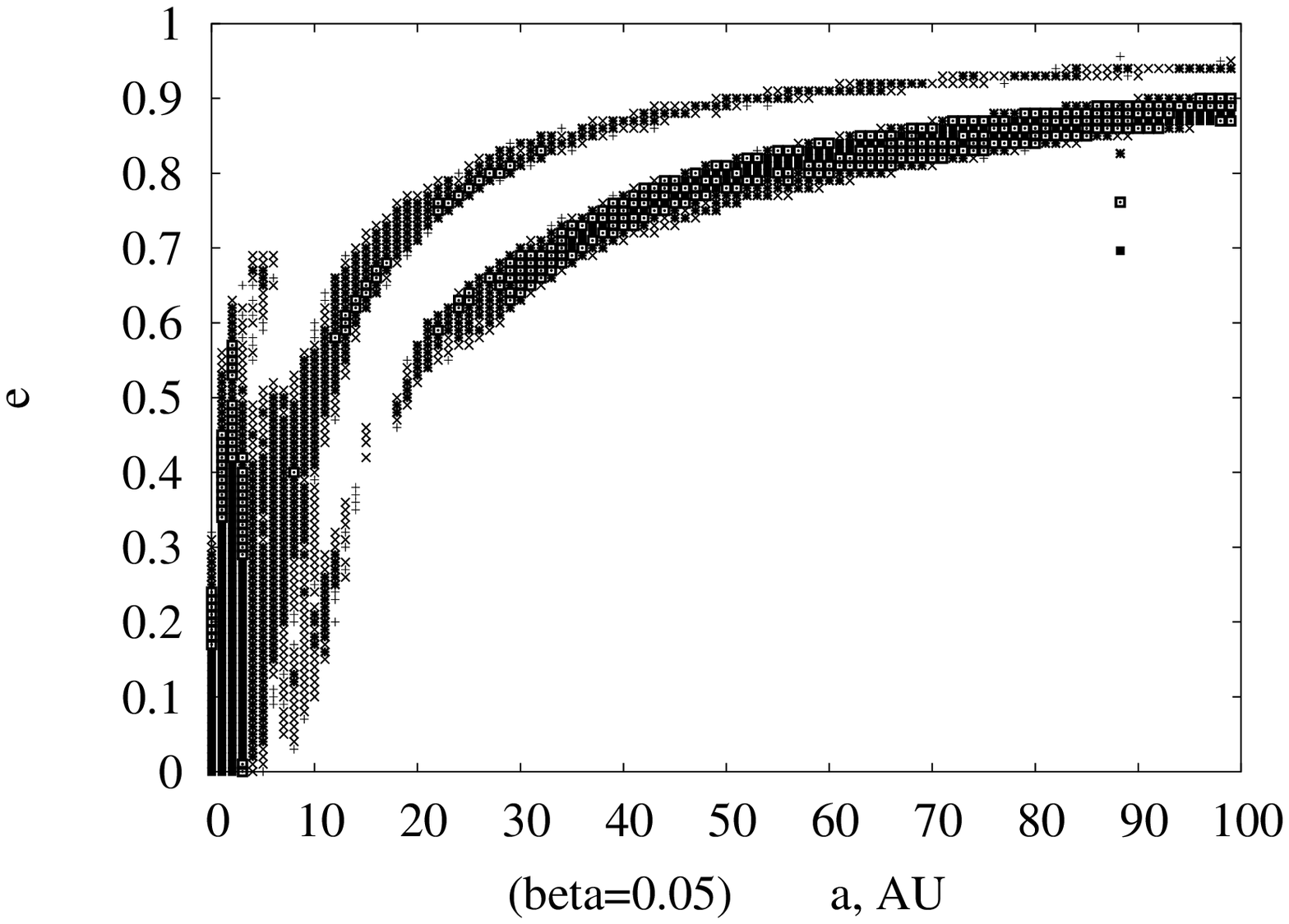}
\includegraphics[width=80mm]{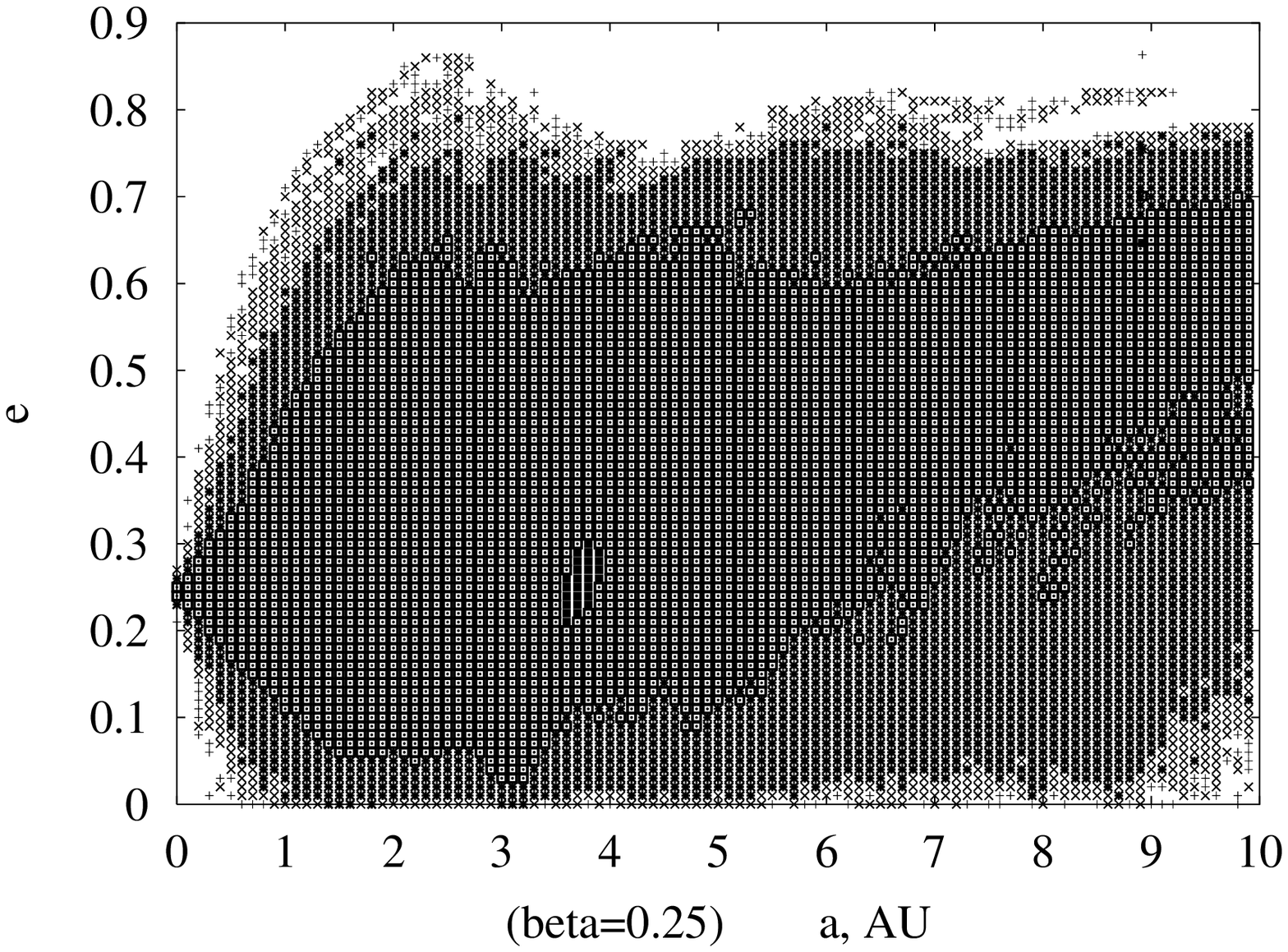}
\includegraphics[width=80mm]{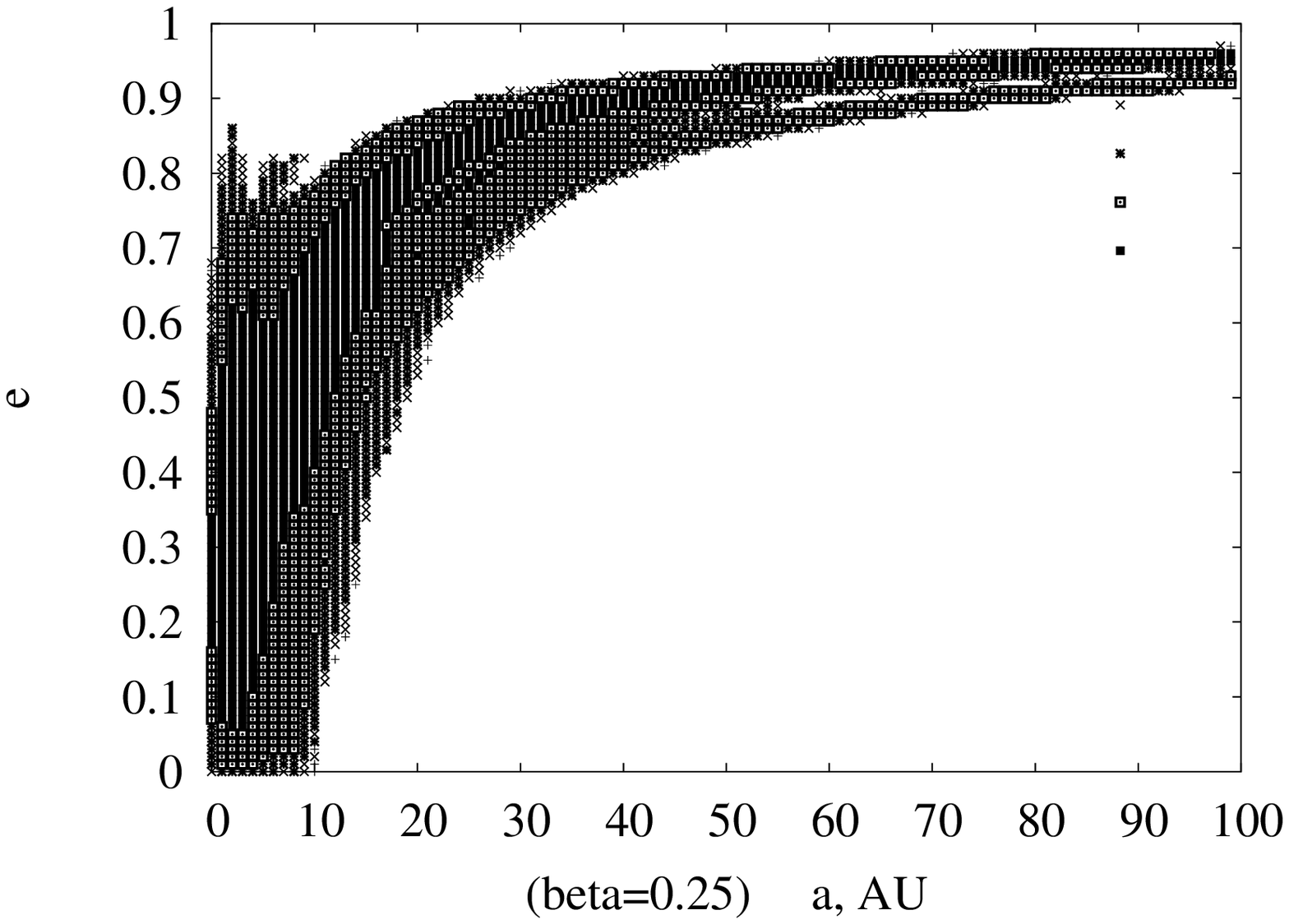}
\includegraphics[width=80mm]{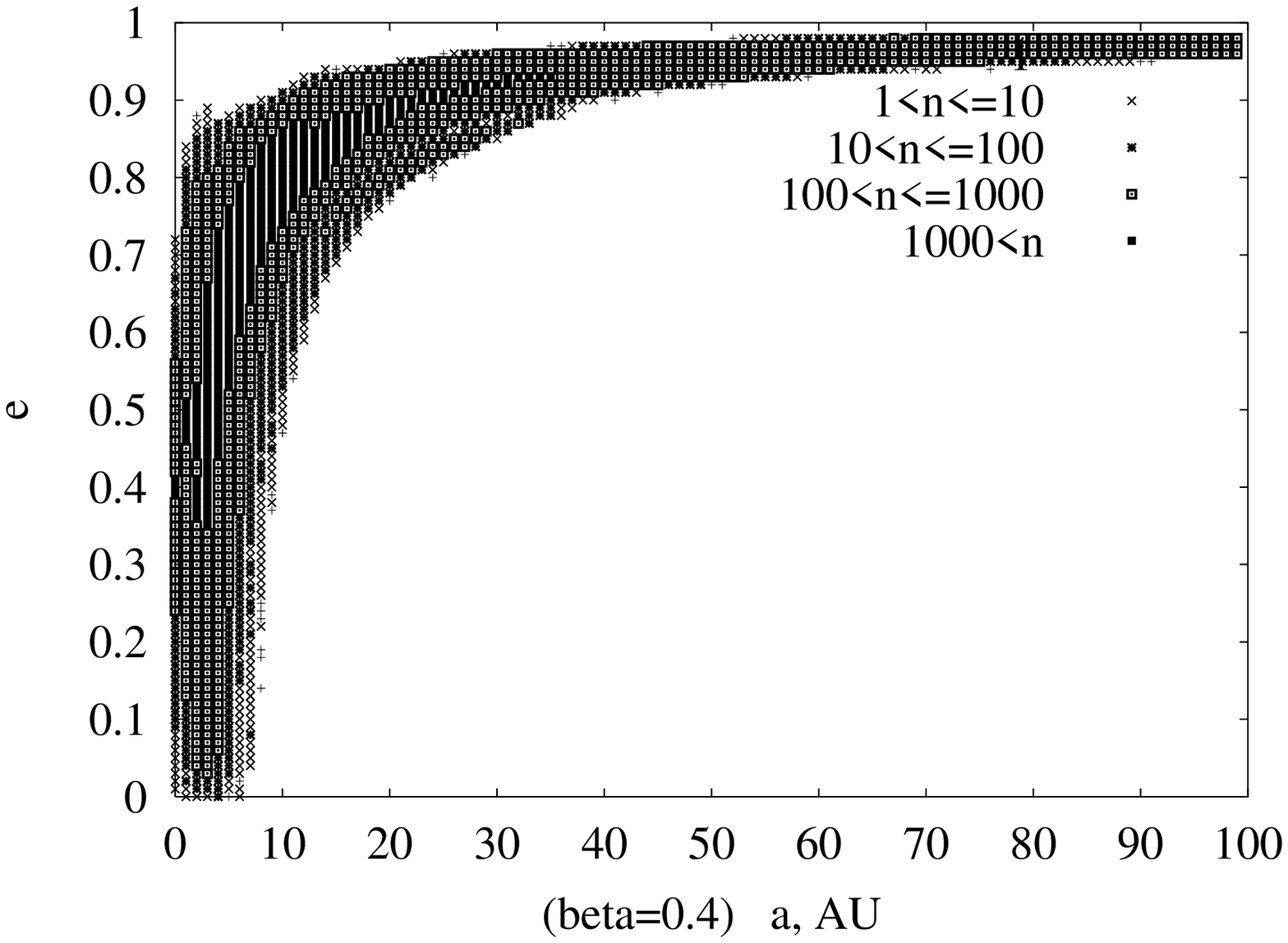}

\caption{Distribution of asteroidal dust particles with semi-major axis and eccentricity
(designations of the number of particles in one bin are 
the same in Figs. 3-7).
For the transfer from rectangular coordinates to
orbital elements we used the same formulas as those for massive bodies.
For Figs. 1-2, 4-8 for calculations of orbital elements we 
added the coefficient (1-$\beta$) to the mass of the Sun,
which is due to the radiation pressure. 
}

\end{figure}%

\begin{figure}
\includegraphics[width=80mm]{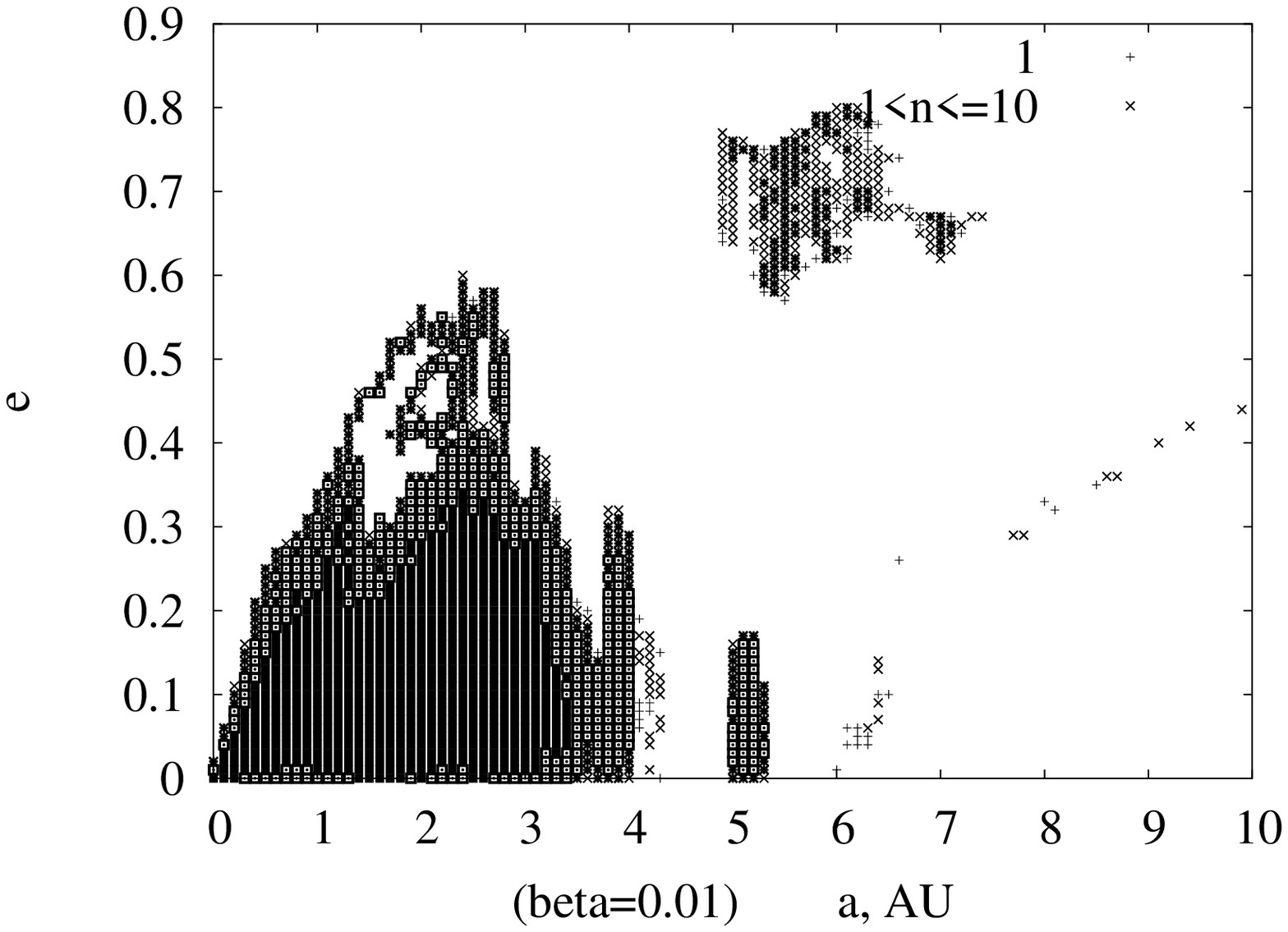}
\includegraphics[width=80mm]{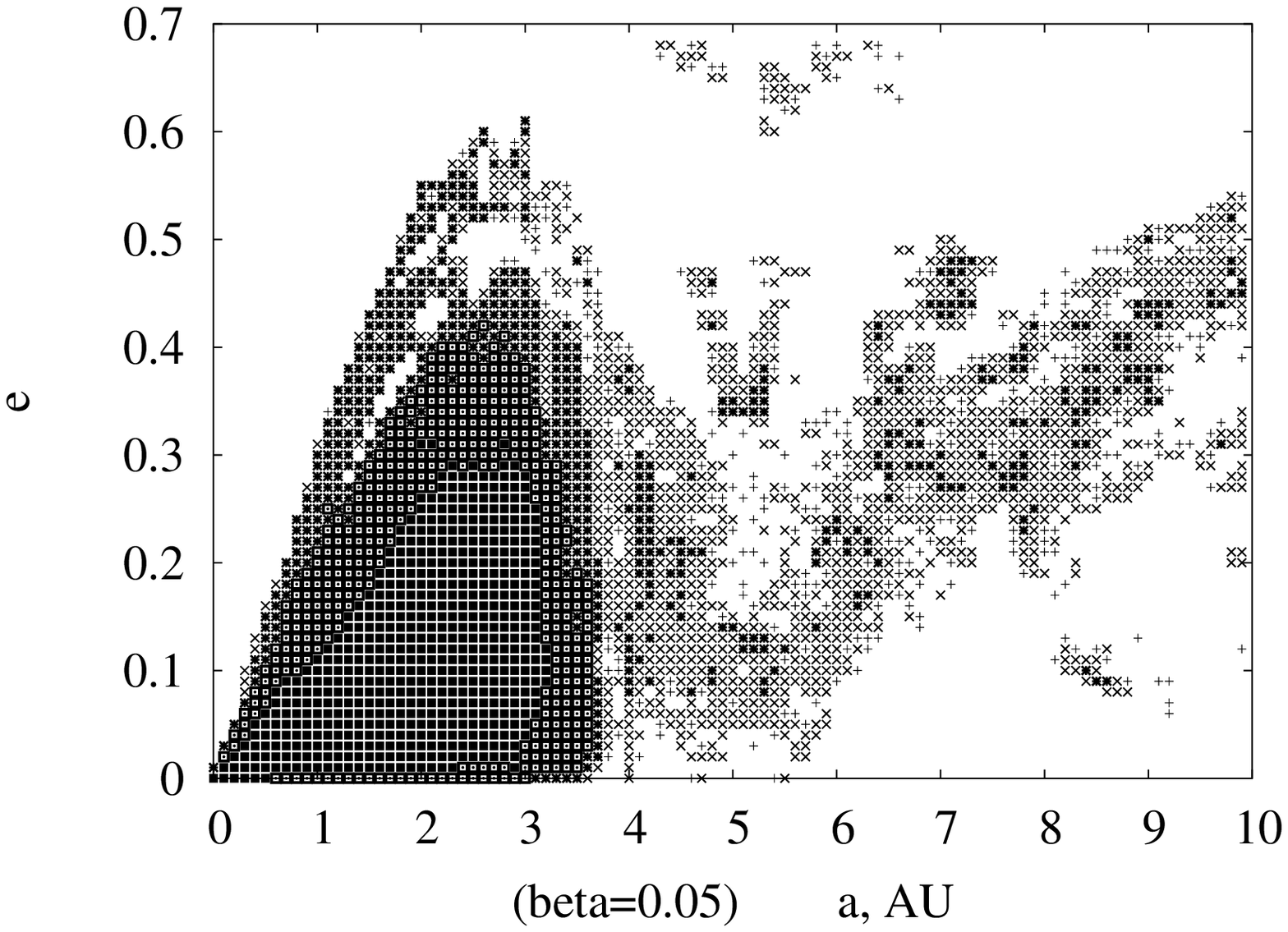}
\includegraphics[width=80mm]{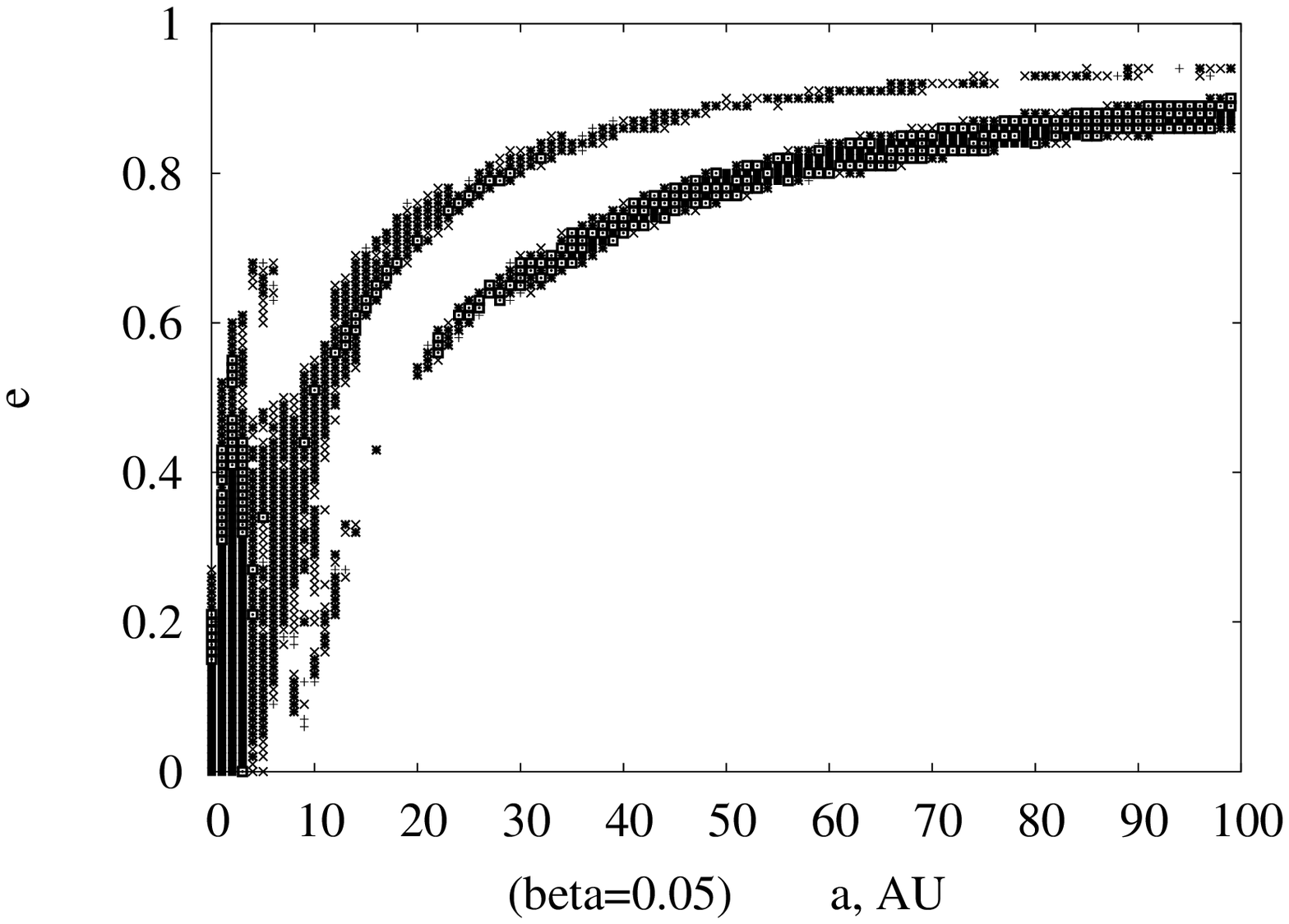}
\includegraphics[width=80mm]{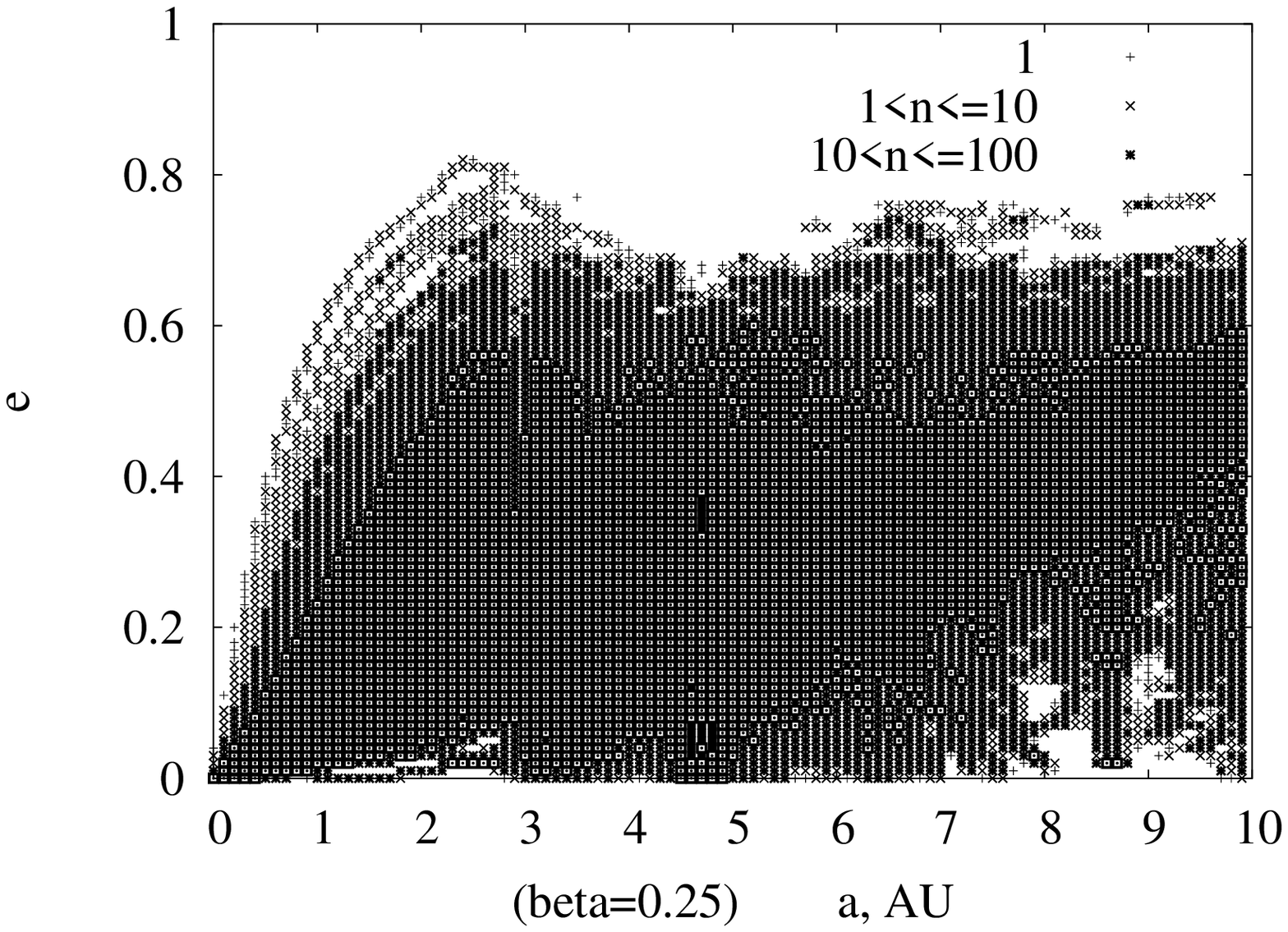}
\includegraphics[width=80mm]{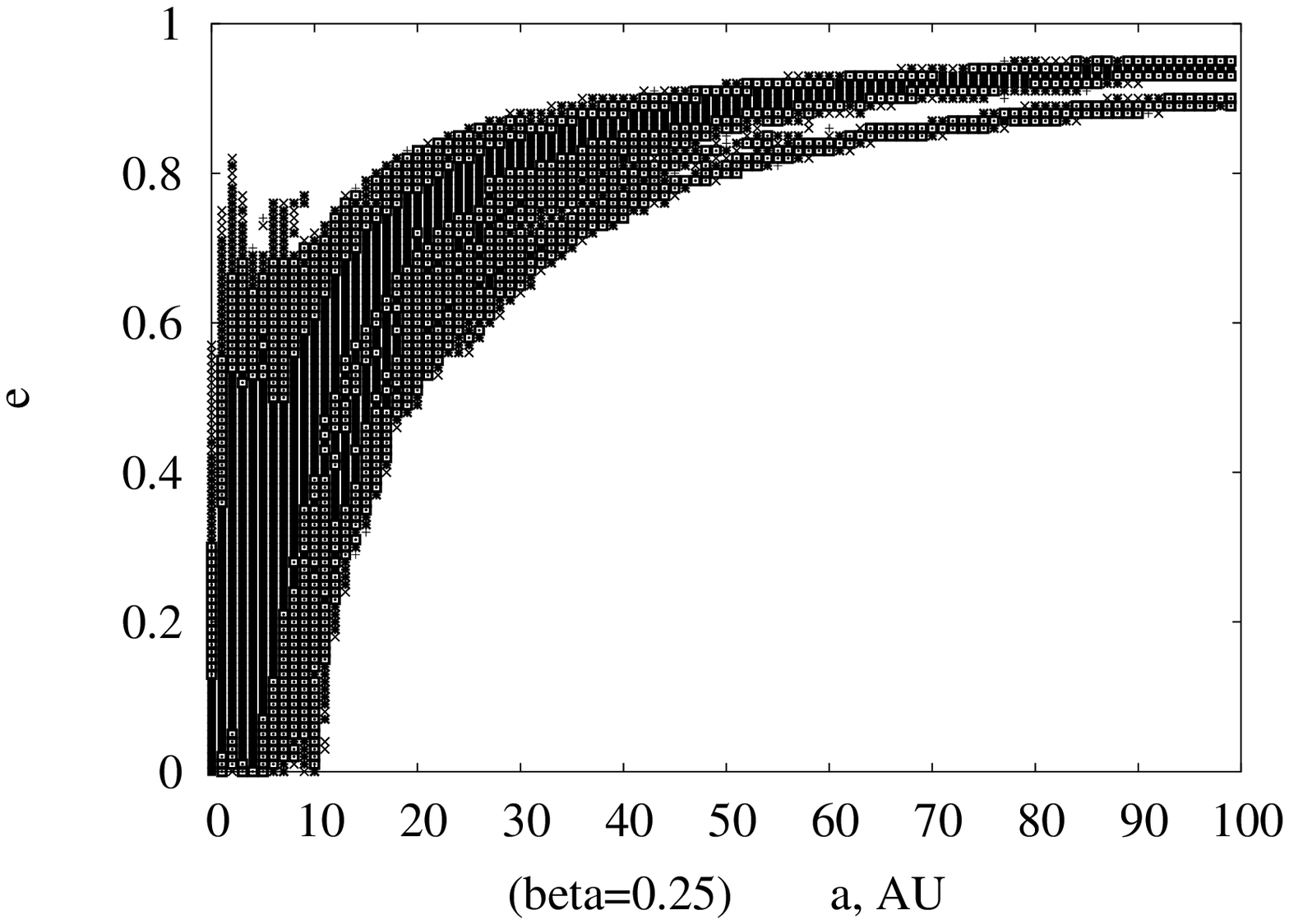}
\includegraphics[width=80mm]{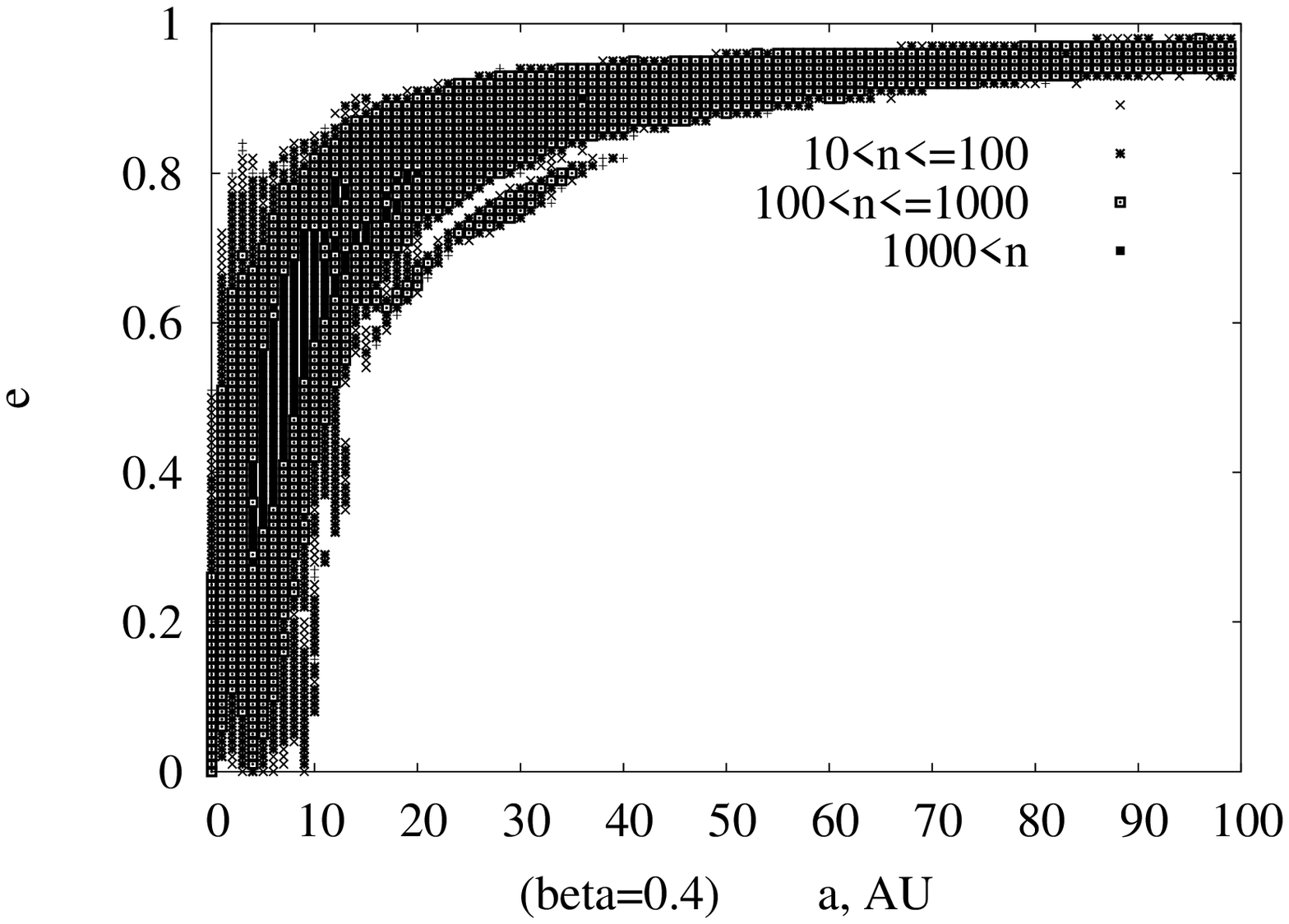}

\caption{Distribution of asteroidal dust particles with semi-major axis and eccentricity
(designations of the number of particles in one bin are 
the same in Figs. 3-7).
}

\end{figure}%

\begin{figure}

\includegraphics[width=80mm]{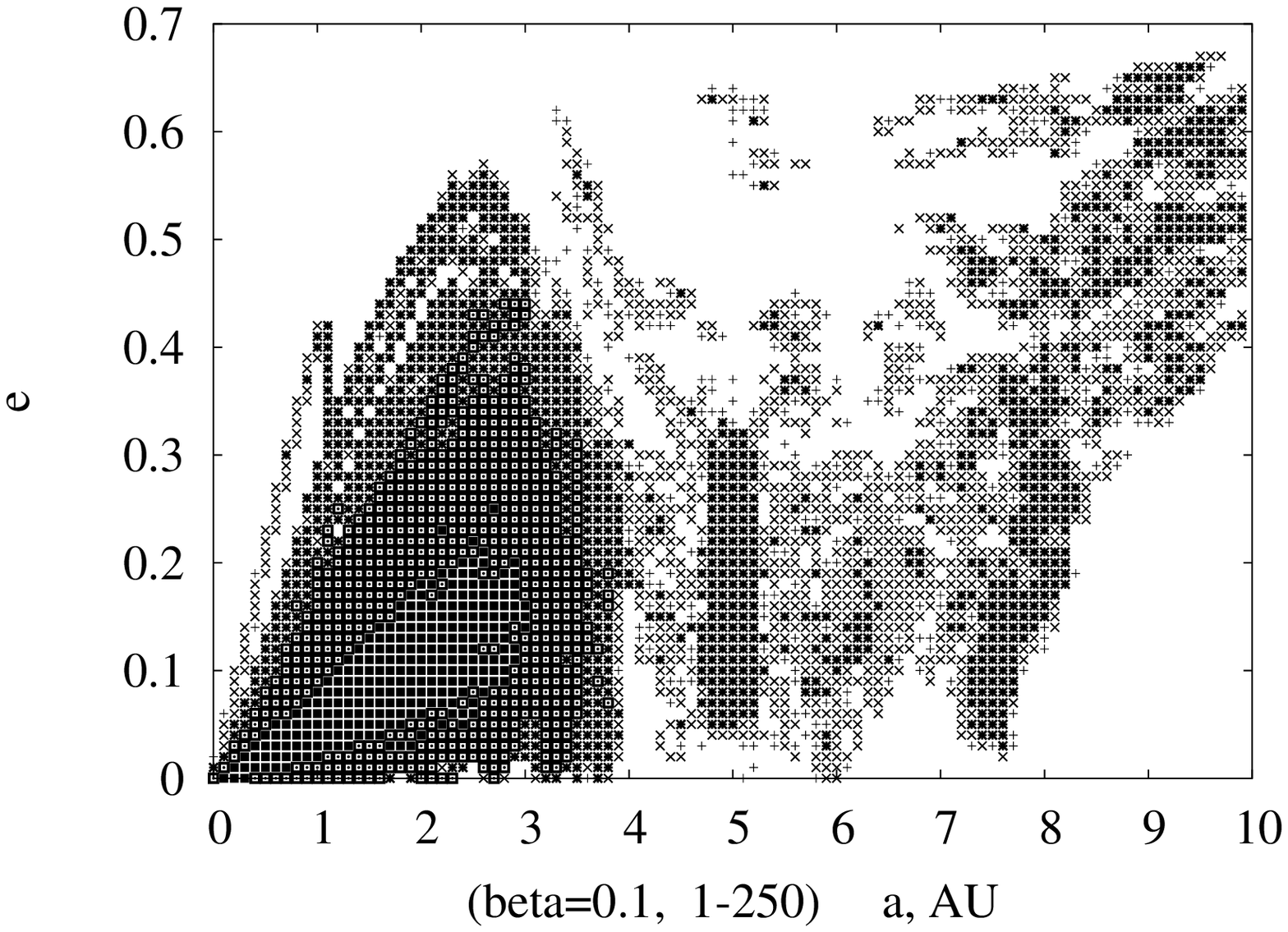}
\includegraphics[width=80mm]{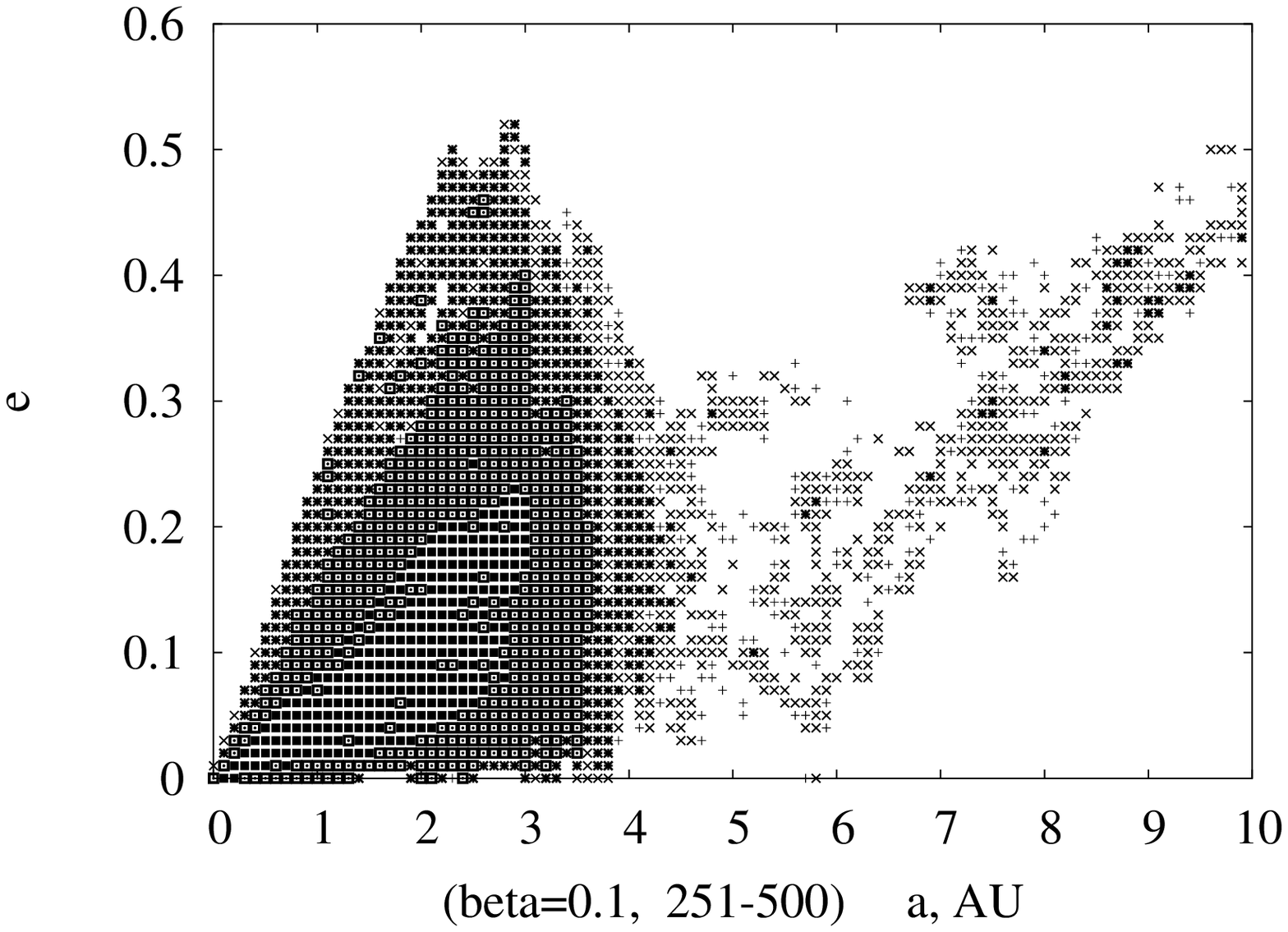}
\includegraphics[width=80mm]{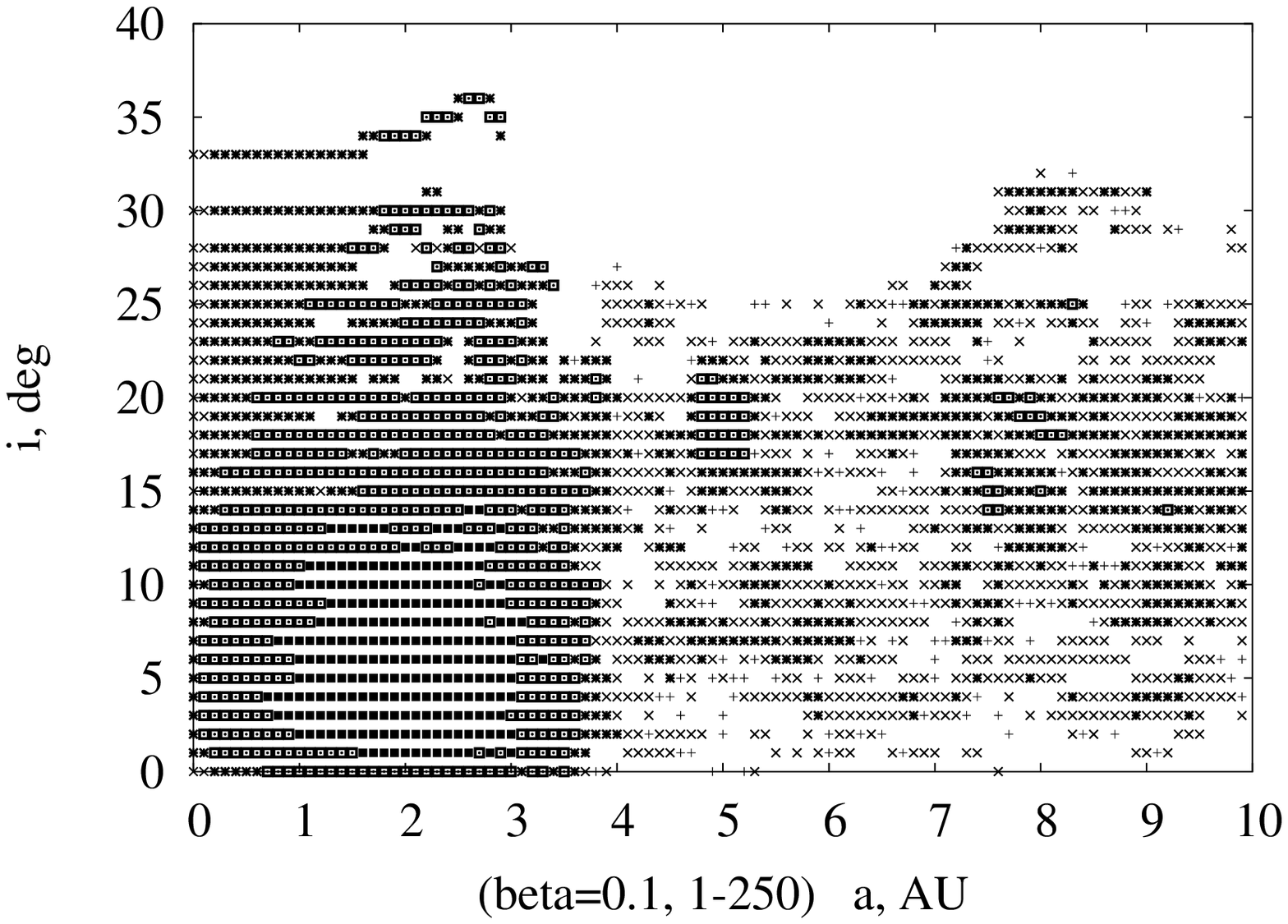}
\includegraphics[width=80mm]{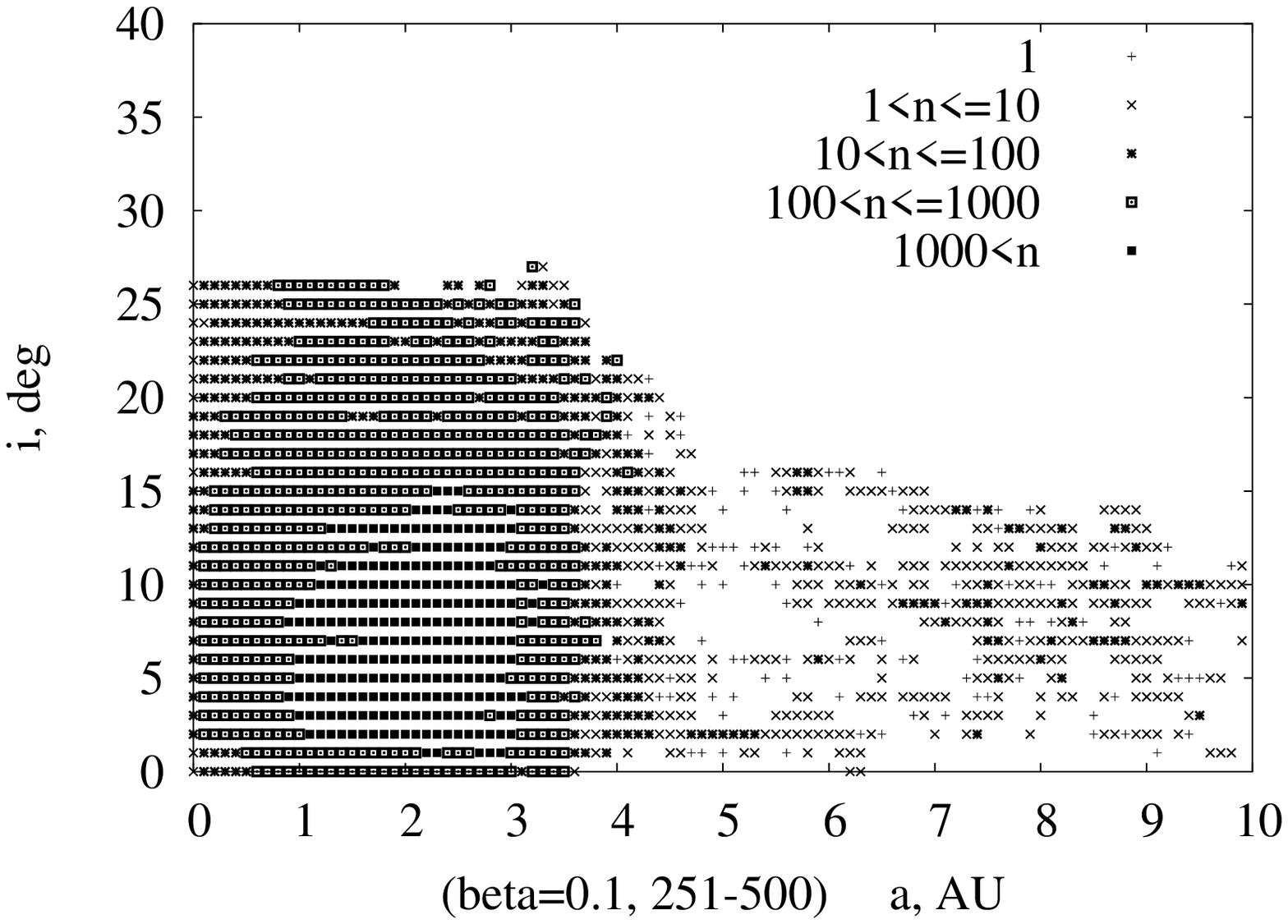}

\caption{Distribution of asteroidal dust particles with semi-major axis and eccentricity,
and with semi-major axis and inclination for $\beta$=0.1. The left figures correspond 
to the runs with initial positions and velocities close to those of the first 250 numbered
main-belt asteroids, and the right figures correspond to the runs 
with initial positions and velocities 
close to those of the asteroids with numbers 251-500 (JDT 2452500.5). 
}

\end{figure}%

\begin{figure}

\includegraphics[width=52mm]{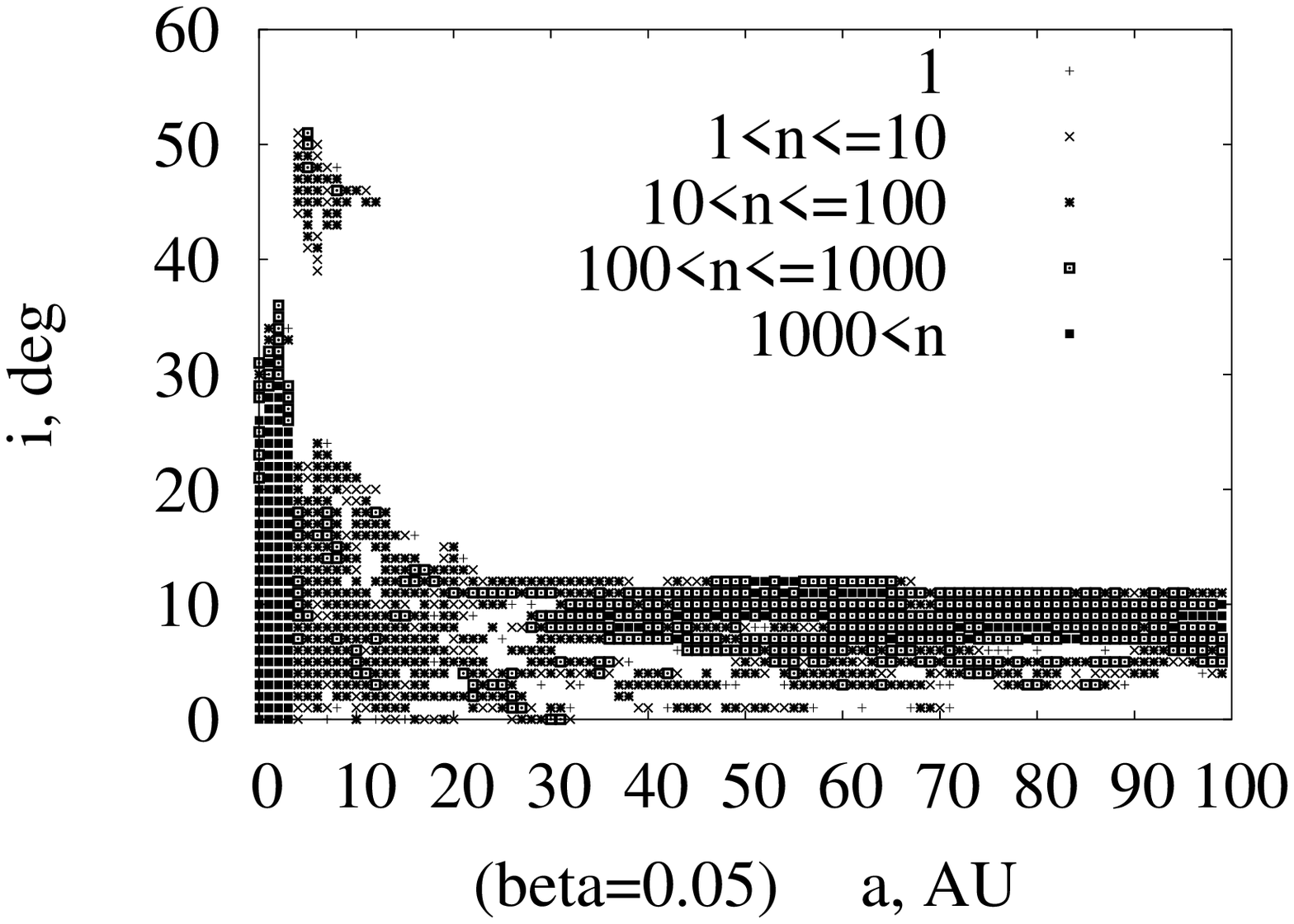}
\includegraphics[width=52mm]{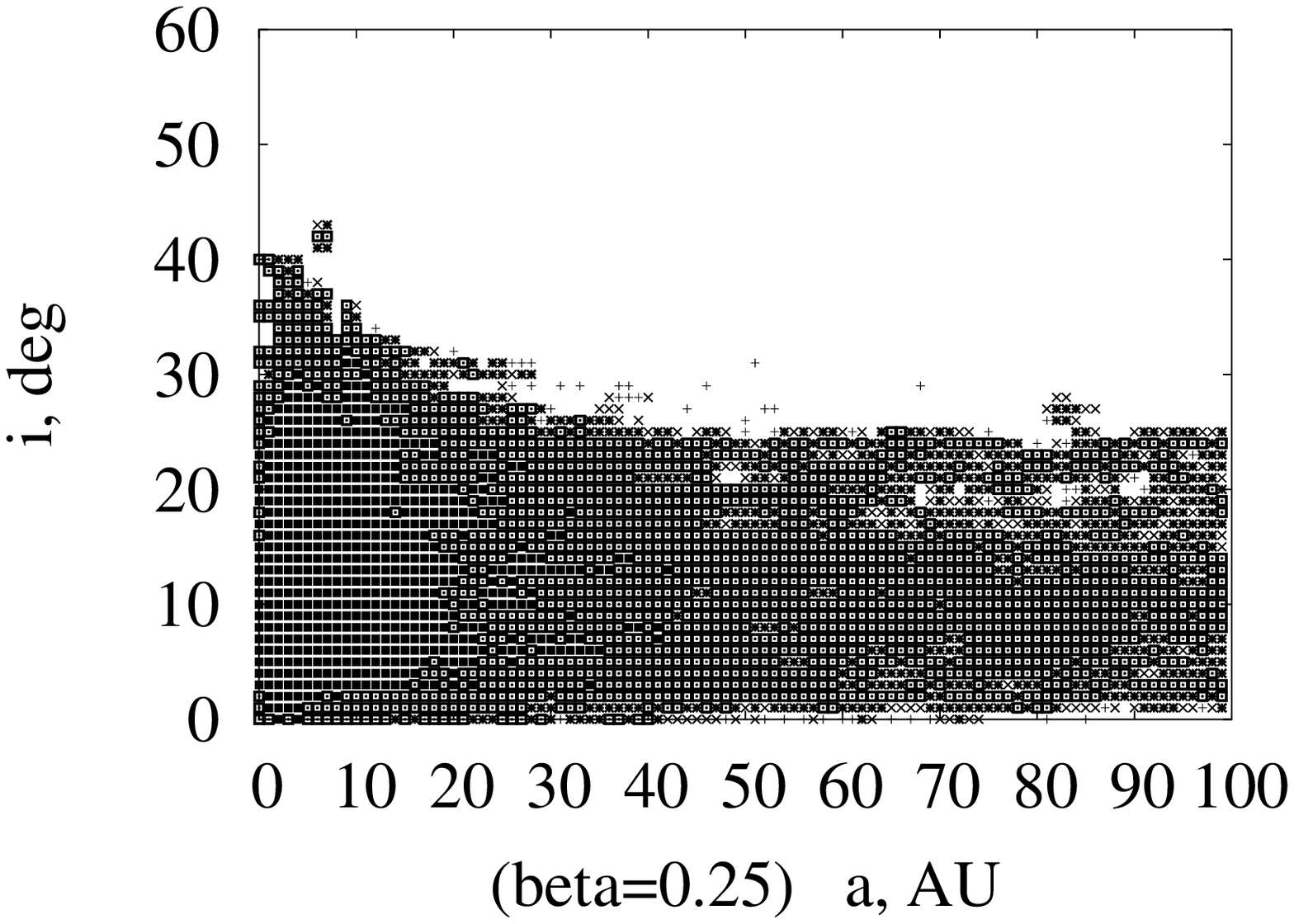}
\includegraphics[width=52mm]{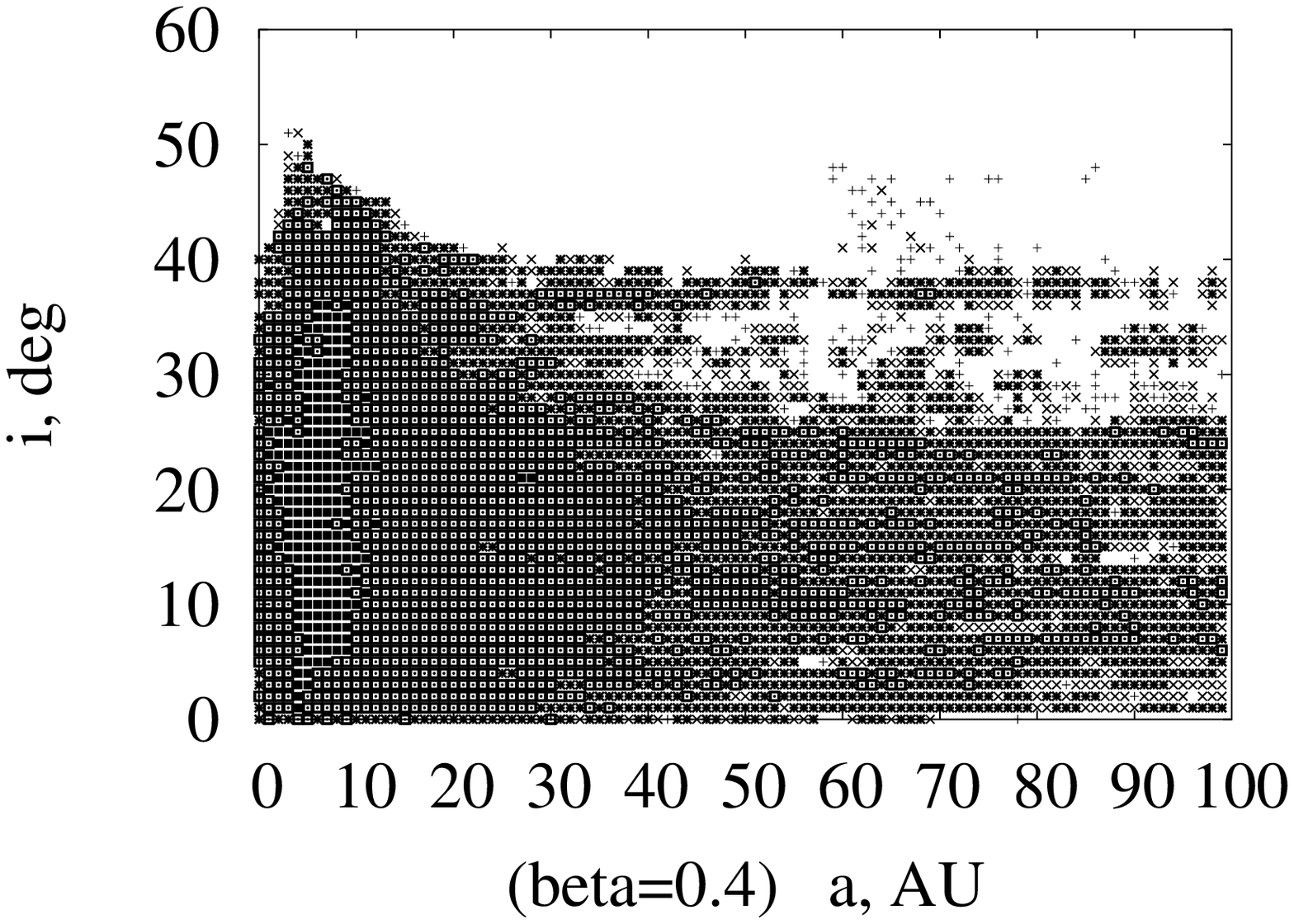}
\caption{Distribution of asteroidal dust particles with semi-major axis and inclination.
}

\end{figure}%

\begin{figure}
\includegraphics[width=80mm]{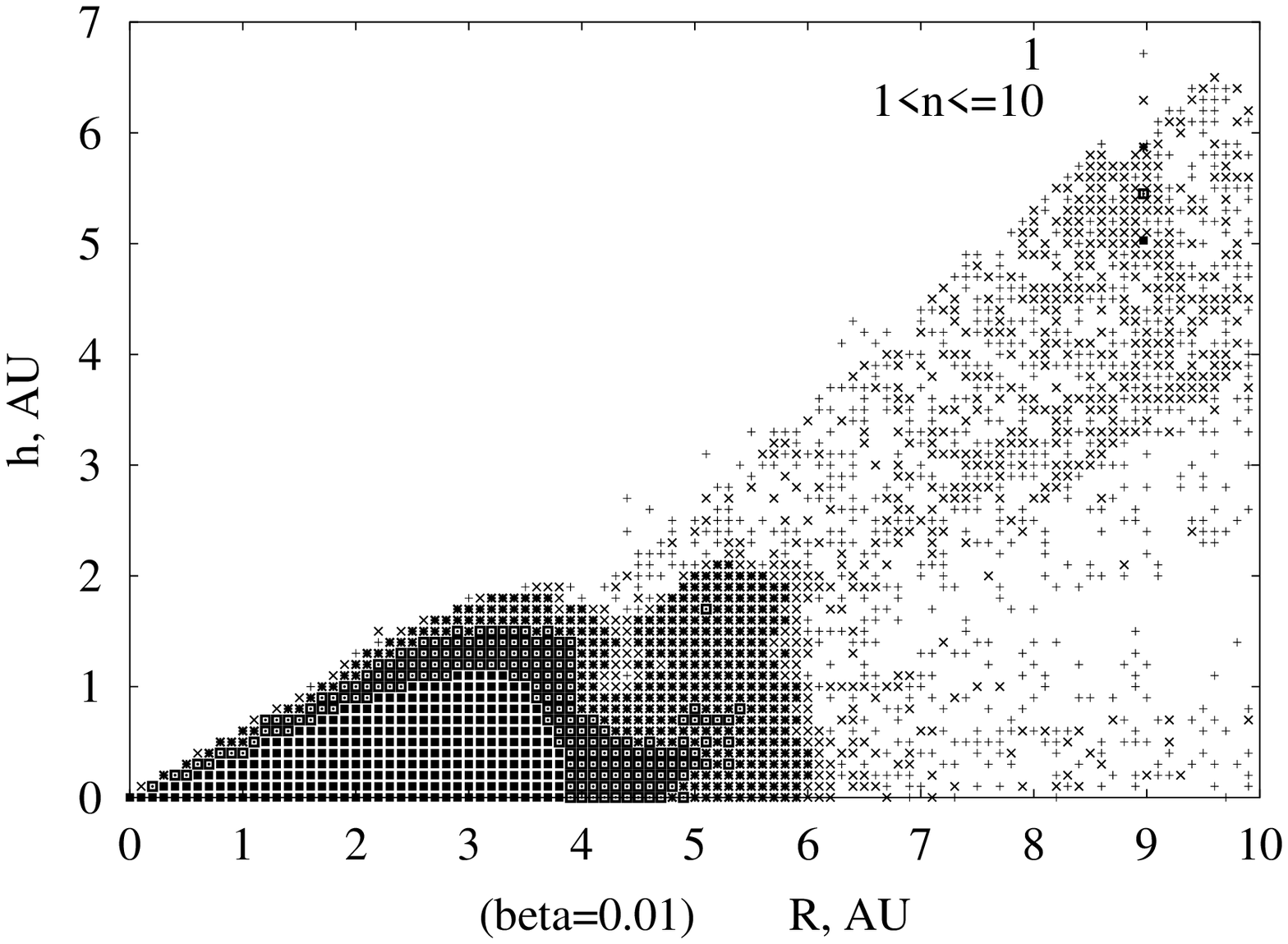}
\includegraphics[width=80mm]{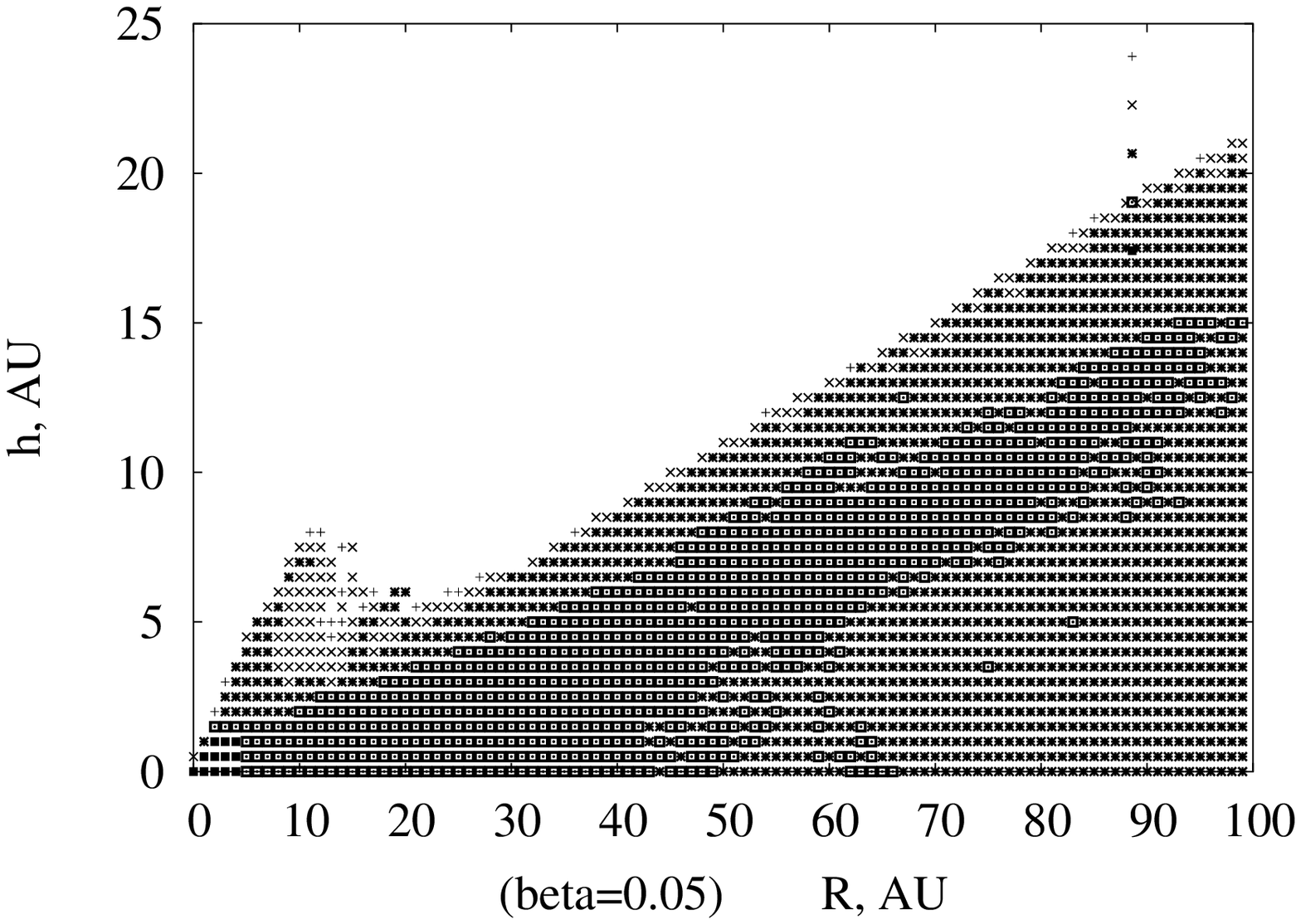}
\includegraphics[width=80mm]{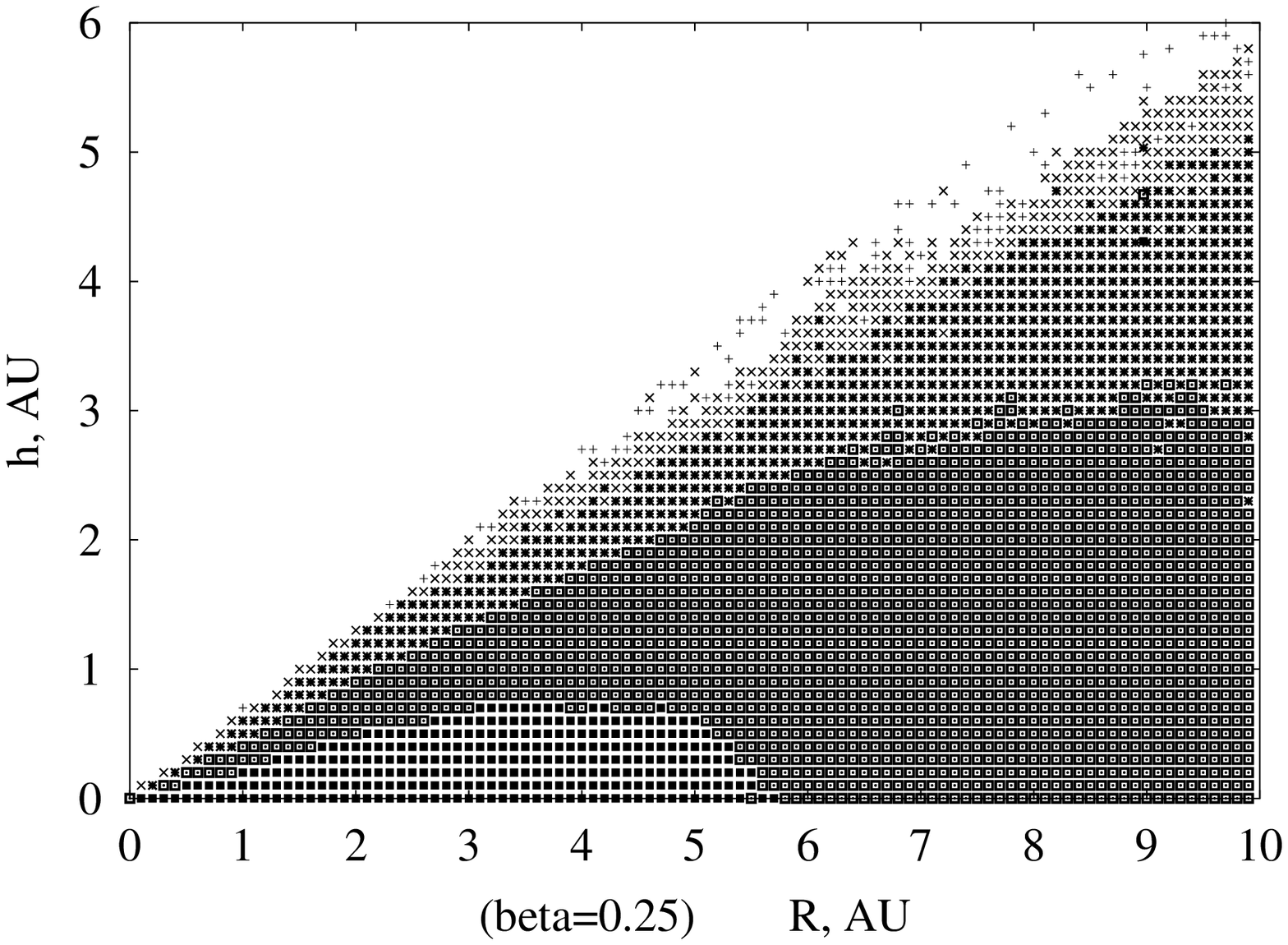}
\includegraphics[width=80mm]{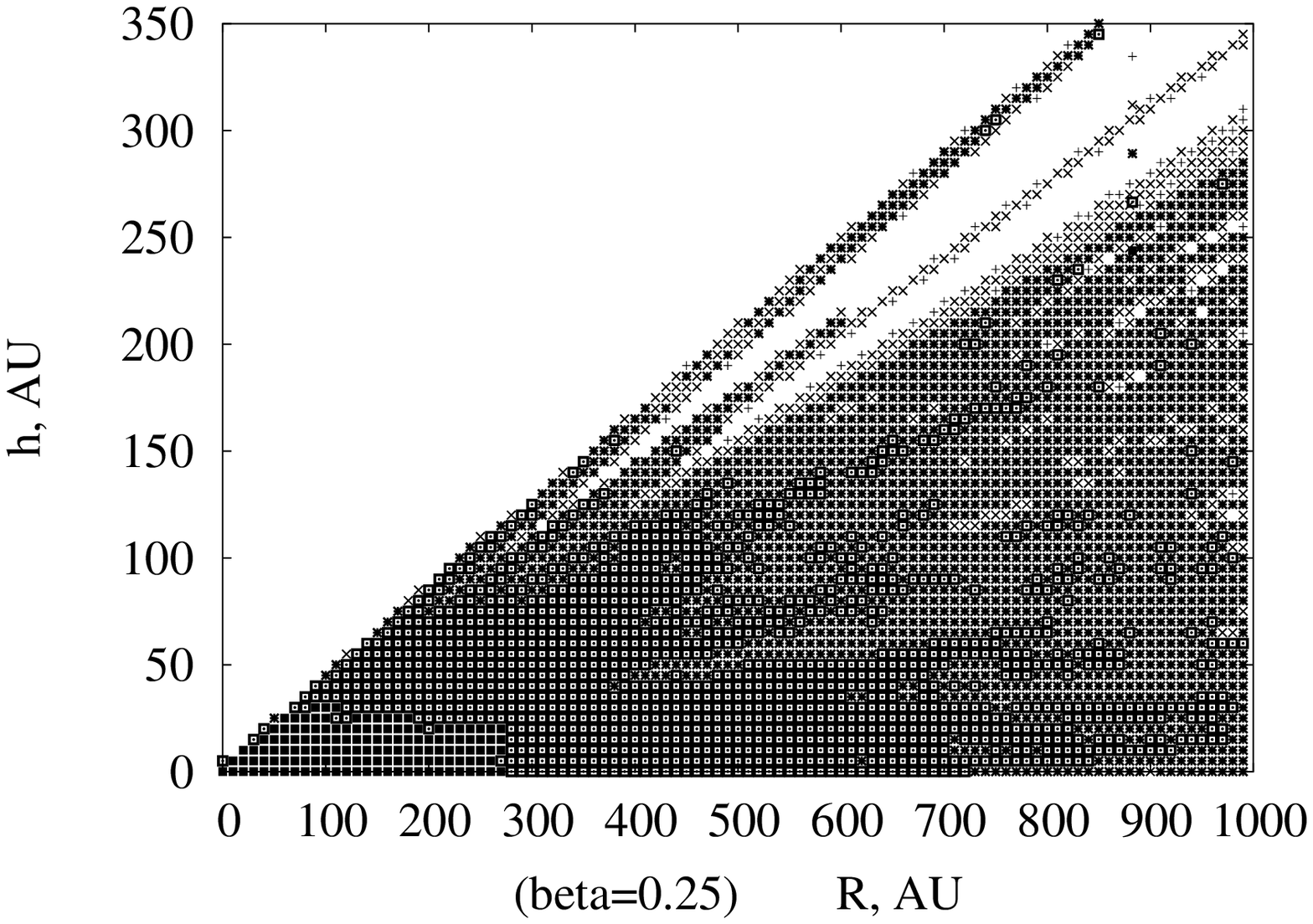}
\includegraphics[width=80mm]{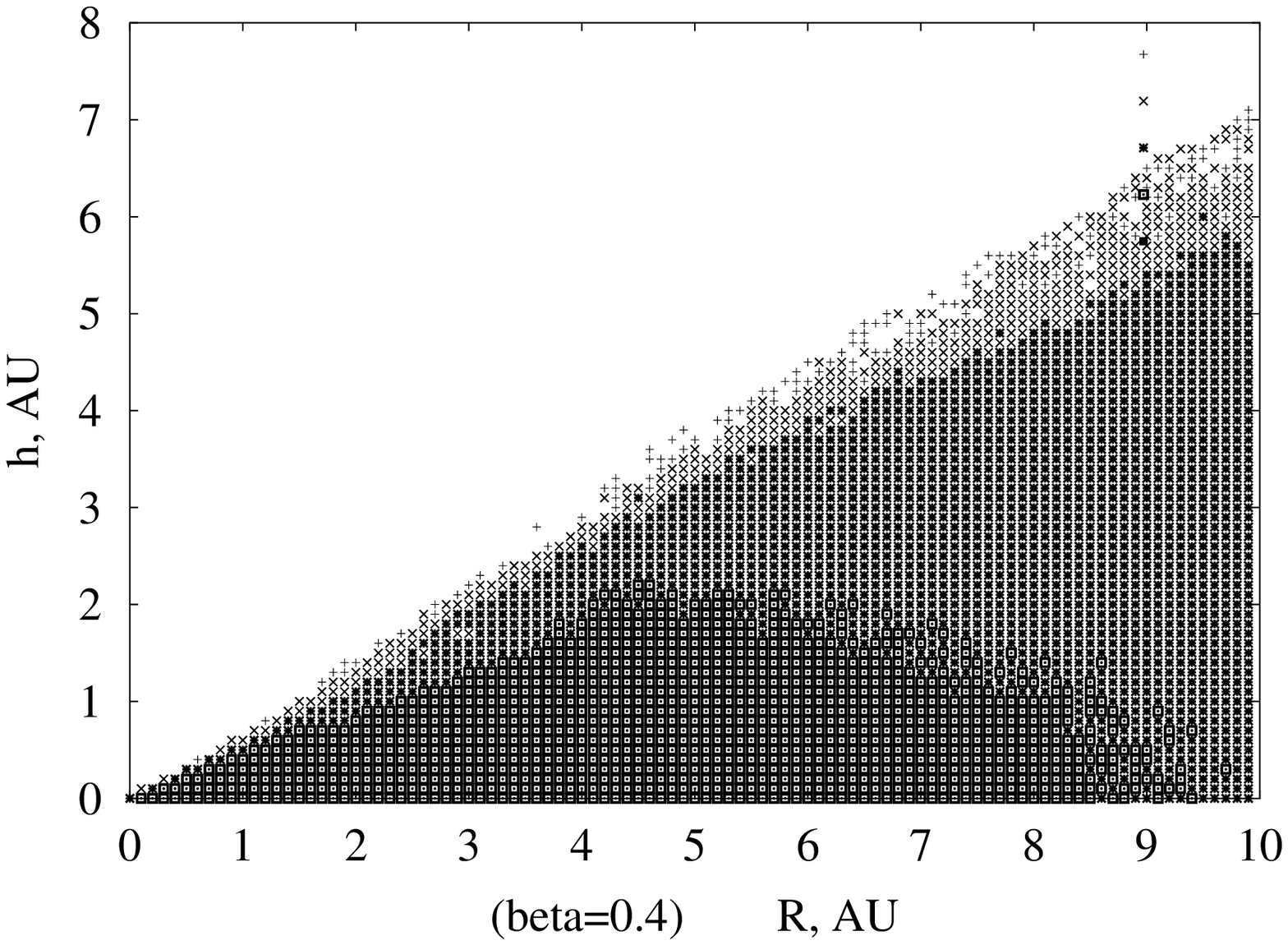}
\includegraphics[width=80mm]{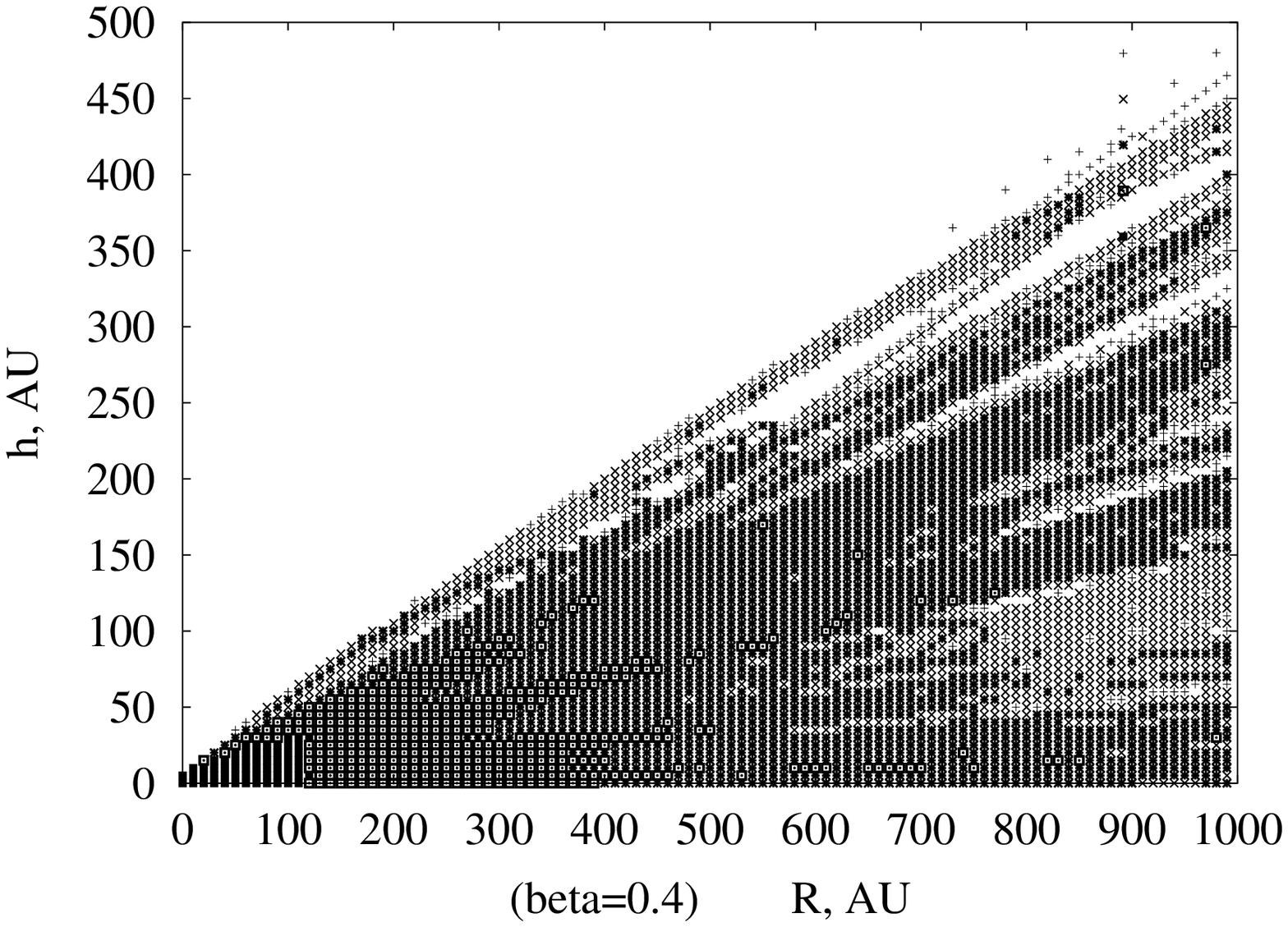}

\caption{Distribution of asteroidal dust particles with distance $R$ from the Sun and 
height $h$ above the initial plane of the Earth's orbit
(designations of the number of particles in one bin are 
the same in Figs. 3-7).
}

\end{figure}%

In Fig. 5 we present the distributions of dust particles in $a$ and $e$, and in $a$ and $i$ for
$\beta$=0.1. The plots on the left were obtained for initial 
positions and eccentricities close to those
of the asteroids with numbers 1-250, and the plots on the right, 
to those of the asteroids with numbers 251-500 
(Figs. 1-4, 6-7 were obtained using all particles).
Overall, the left and the right plots are similar, but for the left plots
particles spent more time outside 5 AU.
At $\beta$=0.05 none of the first 250 particles reached Jupiter's orbit, but
two particles with numbers 361 and 499 migrated to 2000 AU with perihelia
near the orbits of Jupiter and Saturn (Fig. 4).

Usually there are no particles 
with $h/R$$>$0.7 at $R$$<$10 AU, with $h/R$$>$0.25 at $R$$>$20 AU for $\beta$$\le$0.1, and 
with $h/R$$>$0.5 at $R$$>$50 AU for $\beta$$\ge$0.25. 
For $\beta$$\ge$0.25 at 
$R$$<$1000 AU 
the entire region with $h/R$$<$0.3 was not empty (Fig. 7). 

The total time spent by 250 particles in inner-Earth ($Q=a(1+e) < 0.983$ AU), 
Aten ($a<1$ AU, $Q>0.983$ AU), Apollo ($a>1$ AU, $q=a(1-e)<1.017$ AU), and Amor
($a>1$ AU, $1.017<q<1.300$ AU) orbits was 5.6, 1.4, 4.5, and 7.5 Myr at $\beta$=0.01 
and 0.09, 0.08, 0.48, and 0.76 Myr at $\beta$=0.4, respectively. 
The spatial density of a dust cloud and its luminosity
(as seen from outside) were greater for smaller $R$.
For example, depending on $\beta$ they were by a factor of 2.5-8 and 7-25 
(4 and 12-13 at $\beta$$\le$0.05)
greater at 1 AU than at 3 AU for the spatial density and luminosity, respectively. 
This is in accordance with the observations for the inner solar system:  
inversion of zodiacal light observations by the Helios spaceprobe 
revealed a particle density $n(R)\propto R^{-1.3}$,
Pioneer 10 observations between the Earth's orbit and the asteroid belt yielded 
$n(R)\propto R^{-1.5}$, and IRAS observations have yielded $n(R)\propto R^{-1.1}$ [23]. 
The intensity $I$ of zodiacal light falls off with heliocentric distance
$R$ as $I$$\sim$$R^{-\gamma}$, with $\gamma$=2 to 2.5 and beyond about 3 AU
zodiacal light was no longer observable above the background light [17].
As in [2], we approximately defined the brightness of each 
particle  as $R^{-2}$.  Beyond Jupiter's orbit even the number of asteroidal 
particles at some distance $R$ from the Sun is smaller for greater $R$, so 
asteroidal dust particles cannot explain the constant spatial density of dust 
particles at $R$$\sim$3-18 AU. 
At such distances, many of the dust particles could have come from the trans-Neptunian
belt or from passing comets.

\section*{MIGRATION OF KUIPEROIDAL DUST PARTICLES}

We also began a series of runs in which the 
initial positions and velocities of the particles 
were the same as those of the first trans-Neptunian objects (JDT 2452600.5),
and our initial data were different from those in previous papers.
We store orbital elements with a step of 100 yr. 
At the present time these runs have been finished for $\beta$=0.1 and $\beta$=0.2
with $N$=50.

The distributions of dust particles with semi-major axis $a$ and 
eccenticity $e$ or inclination $i$, and 
with distance $R$ from the Sun and 
height $h$ above the initial plane of the Earth's orbit are presented in Fig. 8.
The left plots were obtained for $\beta$=0.1, and the right plots were made for
$\beta$=0.2.  For both values of $\beta$,
particles with $e$$>$0.5 had their perihelia mainly near the semi-major axis of 
the giant planets.
The mean eccentricity $e_m$ of kuiperoidal dust particles
increased with $a$ at
$a$$>$50 AU, and it exceeded 0.5 at	 $a$$>$60 AU.
Bodies that migrated inside Neptune's orbit had $e_m$$<$0.2.
The inclinations were usually less than 35$^\circ$.
Similar figures for intermediate stages of the runs for
kuiperoidal particles were presented in [24] 
(for $\beta$=0.05, $N$=50, and a
time interval of 2 Myr; and  for
$\beta$=0.1, $N$=100, and $t$$\le$3.5 Myr). 

\begin{figure}
\includegraphics[width=81mm]{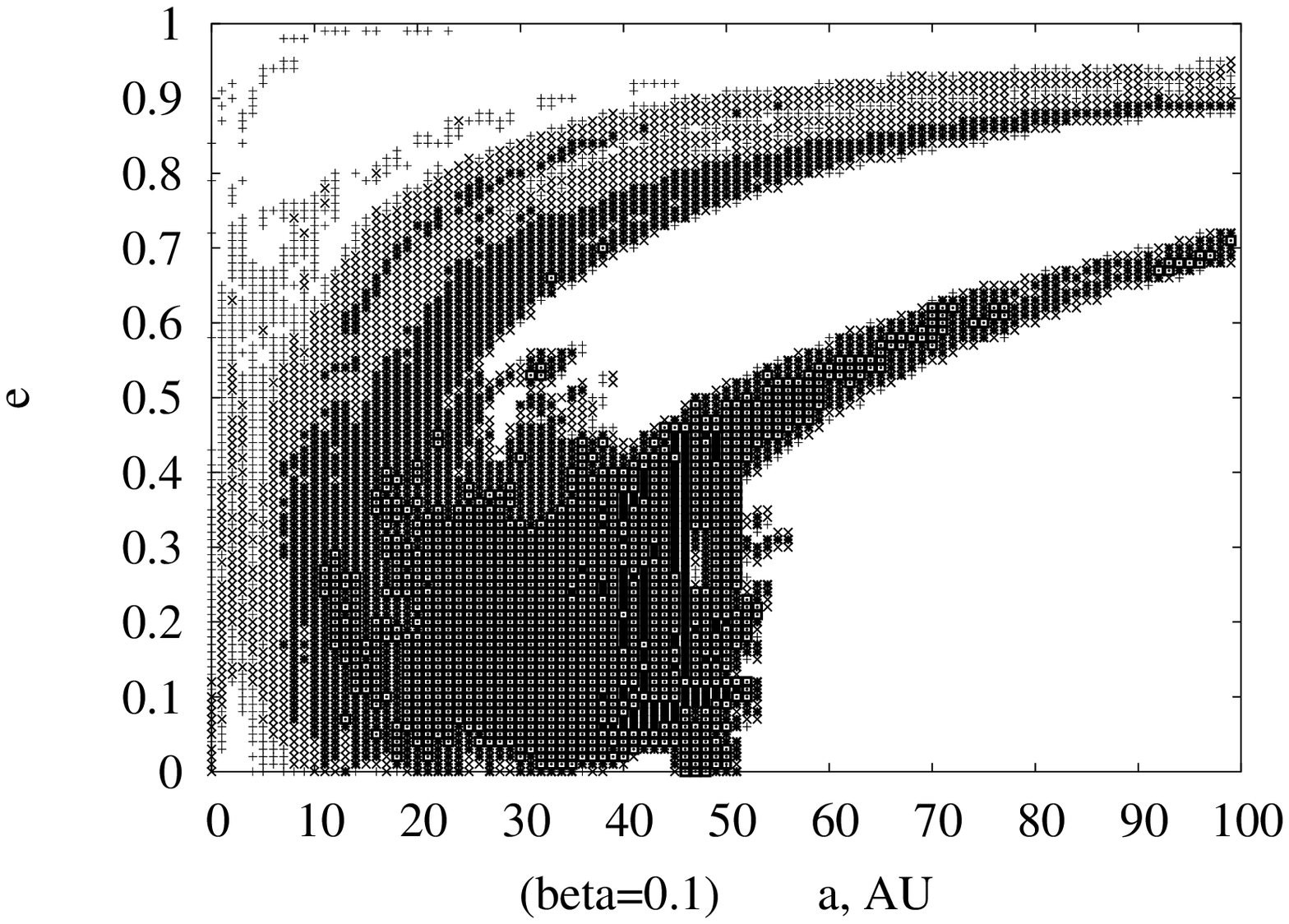}
\includegraphics[width=81mm]{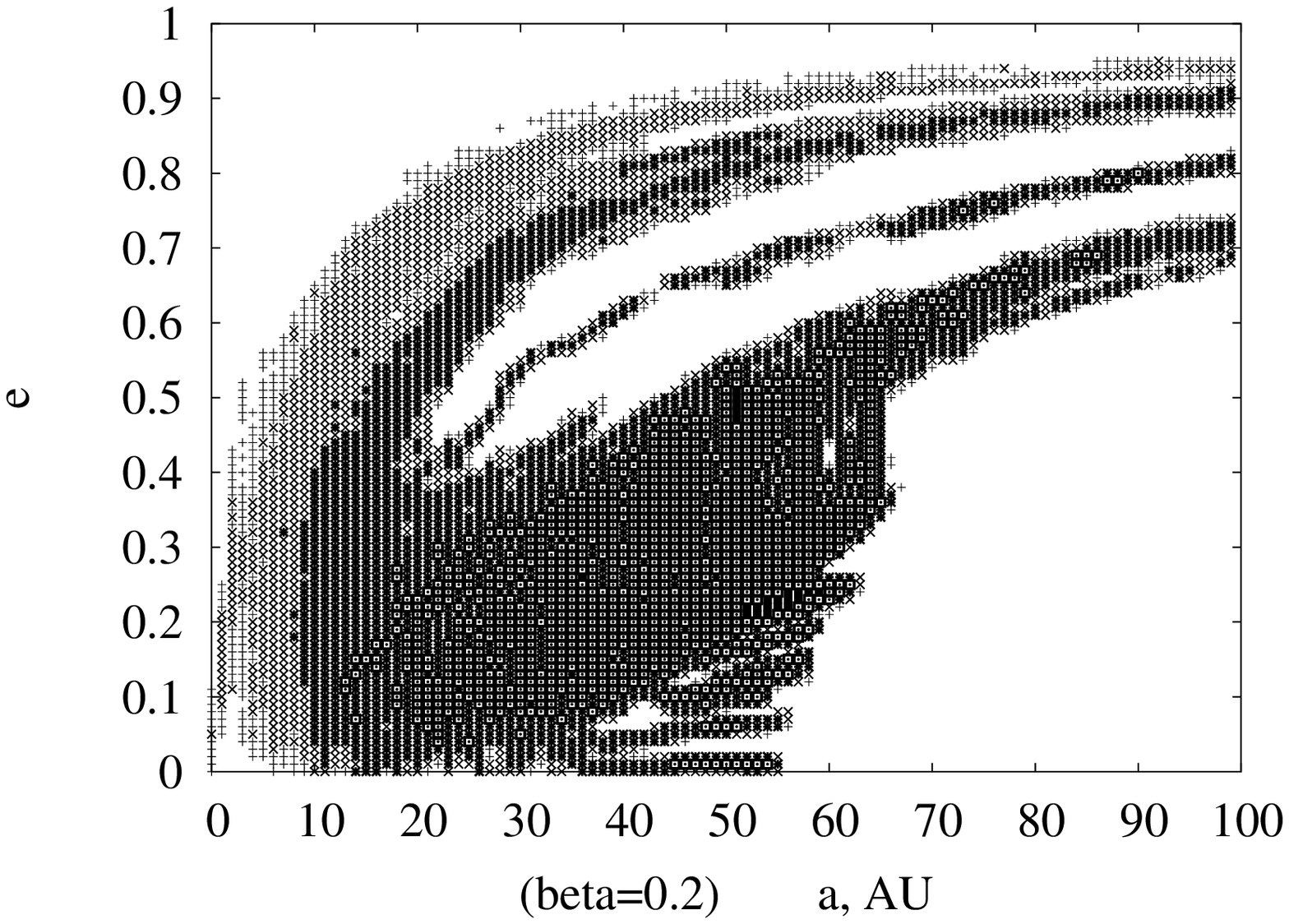}
\includegraphics[width=81mm]{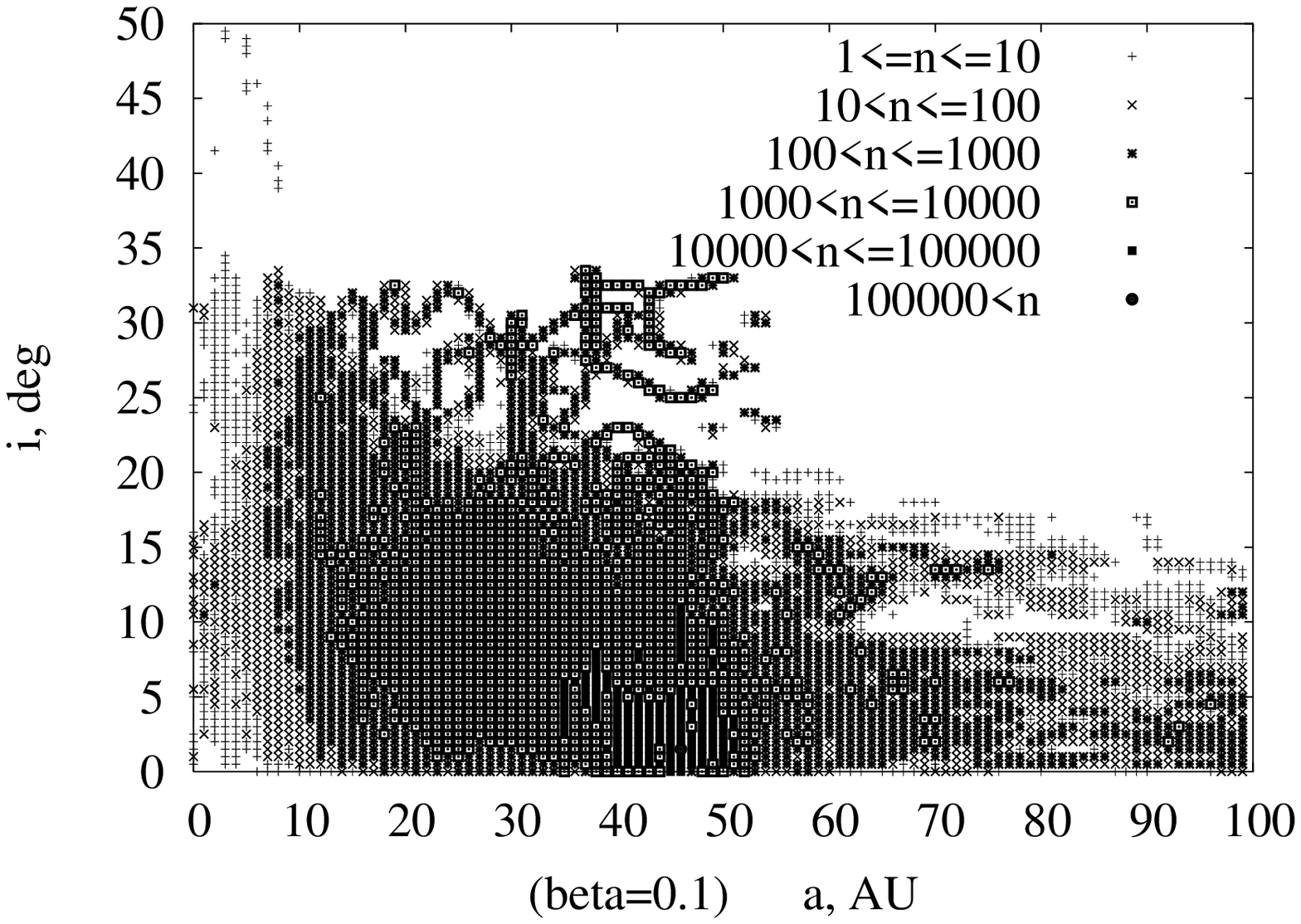} 
\includegraphics[width=81mm]{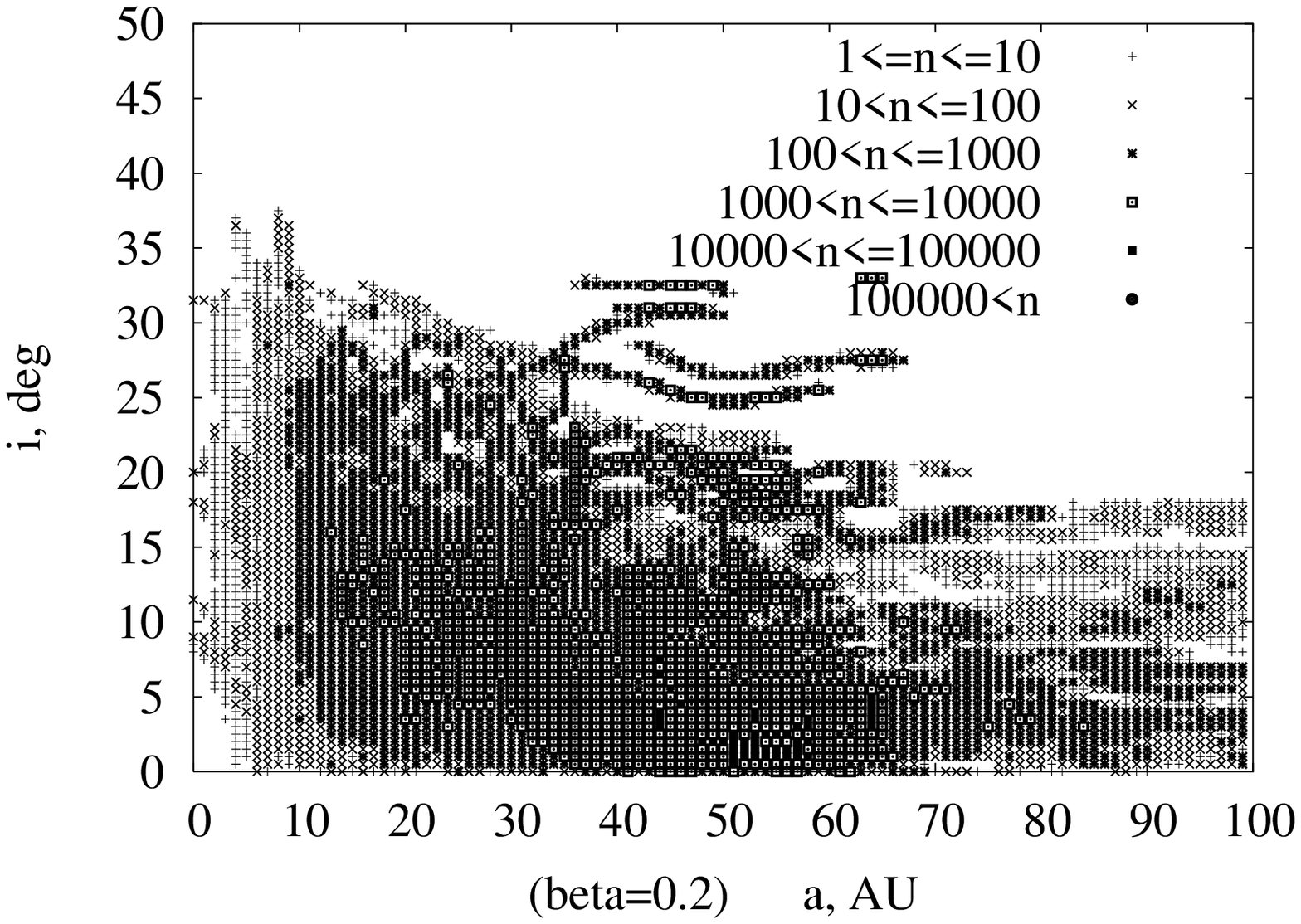} 
\includegraphics[width=81mm]{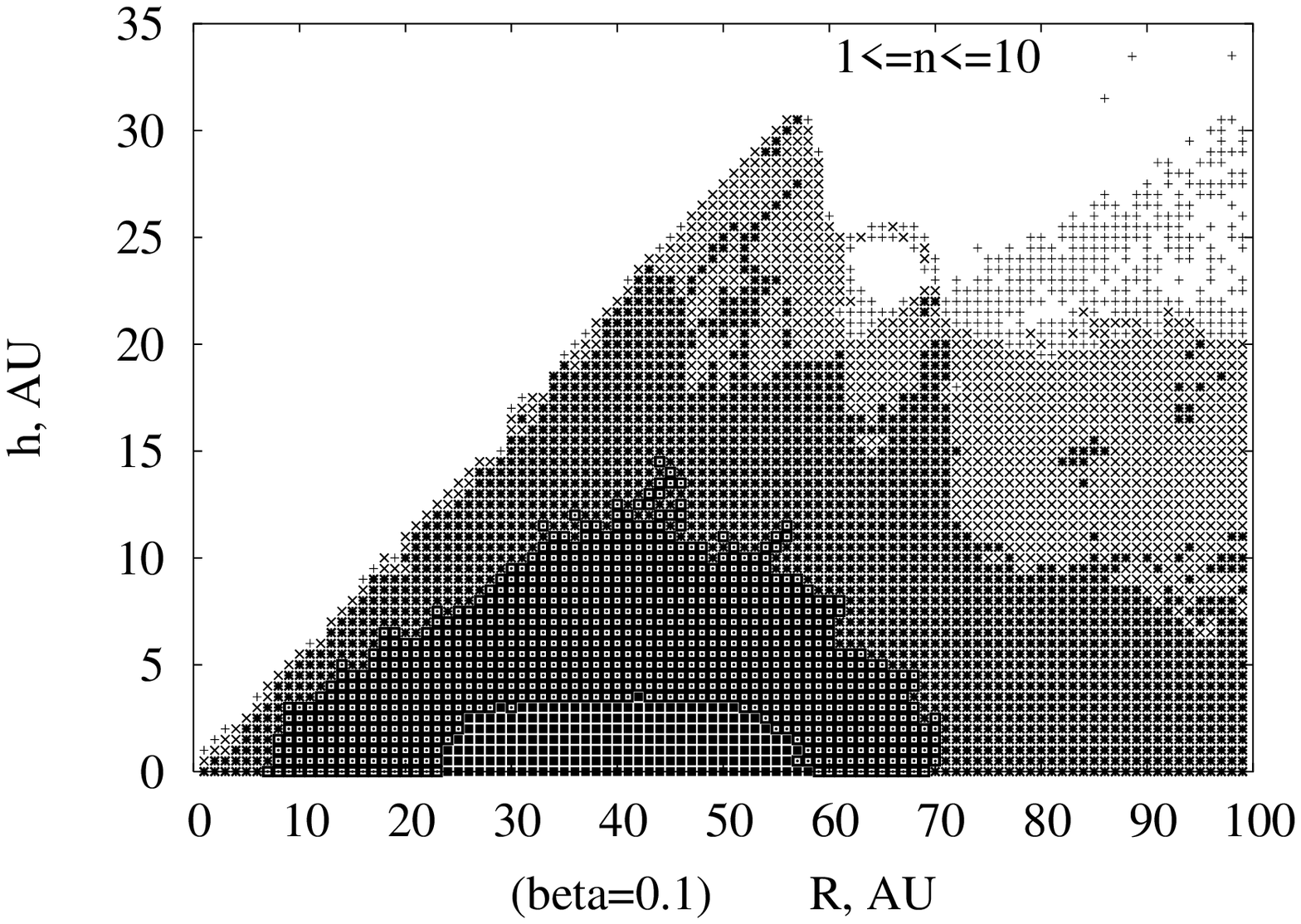}
\includegraphics[width=81mm]{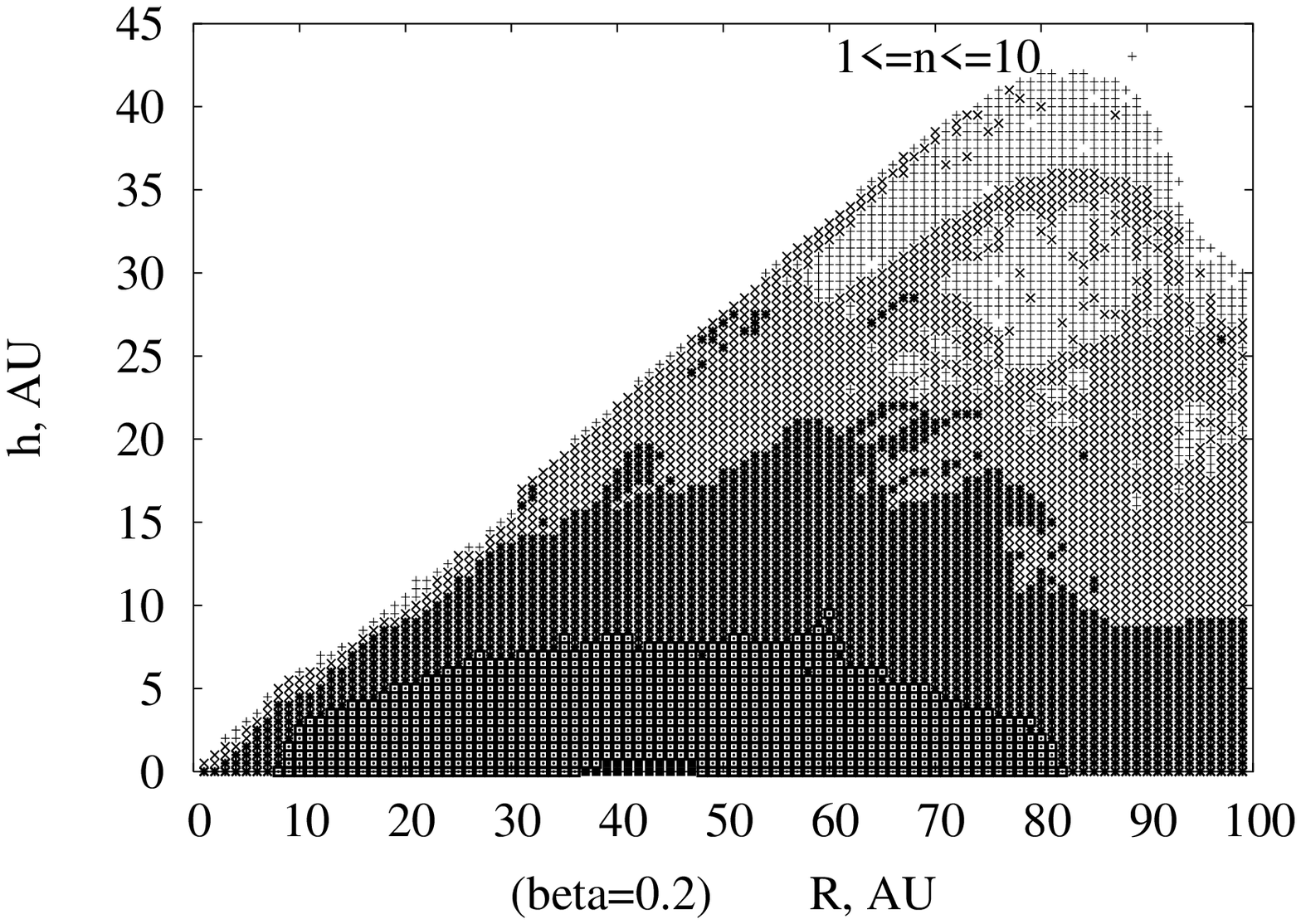}
\caption{Distribution of trans-Neptunian (kuiperoidal) dust particles with semi-major axis $a$ and 
eccenticity $e$ or inclination $i$, and 
with distance $R$ from the Sun and 
height $h$ above the initial plane of the Earth's orbit. 
}
\end{figure}%


At $\beta$=0.4  most particles were outside 50 AU after only 0.02 Myr.
If we consider positions of particles for $t$$\le$0.05 Myr, then 
84\% of them had $R$$>$100 AU. At that time 
the values of $n_R/R$ (surface density), $n_R/R^2$ (spatial density), and
$n_R/R^3$ (similar to luminosity)
were maximum at $\sim$40-55 AU.

In Fig. 8
the values of $h$ were maximum (30 AU) at $R$$\sim$57 AU for $\beta$=0.1,
and were maximum (42 AU) at $R$$\sim$80 AU for $\beta$=0.2.
We usually obtained $h/R$$<$0.5, and 
in the zone of the Edgeworth--Kuiper belt
the ratio of the number of particles at $h$=$kR$ 
dropped usually to 10\% of the number at $h$=0  at $k$=0.1, but sometimes at $k$=0.25. 

\begin{table}
\caption{Values of $T$, $T_J$, $T_S^{min}$, $T_S^{max}$, $T_{2000}^{min}$,
$T_{2000}^{max}$ (in Kyr), $P_r$, and $P_{Sun}$ 
obtained for kuiperoidal dust particles at $\beta$=0.1 and $\beta$=0.2
(Venus=V, Earth=E, Mars=M)
}

$ \begin{array}{lcccccccccccc} 

\hline	

  & & $V$ & $V$ & $E$ & $E$ & $M$ & $M$ & && && \\


\beta & P_{Sun}& P_r & T & P_r & T &  P_r & T & T_J & T_S^{min}&T_S^{max}&
T_{2000}^{min}&T_{2000}^{max} \\

\hline
0.1 & 0.2&76.2 & 0.75 & 35.2  & 1.42 & 2.74 & 2.79 & 47.2 &3659&  17439&3730&53949\\
0.2  & 0.12& 182  & 0.22  & 150  & 0.46 & 13.3 & 1.2 & 59.6 &5237&10789&2490& 26382 \\

\hline
\end{array} $ 
\end{table}	

Table 2 is similar to Table 1, but was obtained for kuiperoidal dust particles.
The fraction $P_{Sun}$ of kuiperoidal particles collided with the Sun was smaller
by a factor of 6 than that of asteroidal particles for the same $\beta$,
but the difference in collision probabilities with Earth and Venus 
for asteroidal and kuiperoidal particles was smaller than 6.
This is due to that the mean eccentricities of particles near these planets
(especially, at $\beta$=0.2) were smaller for kuiperoidal particles than
those for asteroidal particles.

The trans-Neptunian belt is considered to be the main source of Jupiter-family comets, 
which can produce much trans-Neptunian dust just
inside Jupiter's orbit. Some of these comets can reach typical near-Earth objects' 
orbits [19]. The total mass of comets inside Jupiter's orbit
is much smaller than the total mass of asteroids, but a comet produces more dust 
per unit minor body mass than an asteroid.

\section*{MIGRATION OF DUST PARTICLES FROM COMET ENCKE}

We also investigated the migration of cometary dust particles 
for $\beta$ equal to 0.002, 0.004, 0.01, 0.05, 0.1, 0.2, and 0.4. For silicate particles 
such values of $\beta$ correspond to diameters equal to about 200, 100, 40, 9, 4, 
2, and 1 microns, respectively. 
The initial positions and velocities of the particles were the same as 
those of Comet 2P Encke. 
We considered particles starting near perihelion 
(runs denoted as $\Delta t_o =0$), 
near aphelion ($\Delta t_o =0.5$), and when the comet had orbited for $P_a/4$ 
after perihelion passage, where $P_a$ is the period of the comet
(such runs are denoted as $\Delta t_o =0.25$).
Variations in time $\tau$ when perihelion was passed was varied 
with a step 
$0.1$ day for series 'S' and with a step 1 day
for series 'L'.
For each $\beta$ we considered $N$=101 particles for "S" runs 
and 150 particles for "L" runs. 


The results obtained are presented in Table 3. 
Again, $T$ is the mean time during which the perihelion distance $q$
was less than the semi-major axis of a given planet. In contrast to the
asteroidal dust particles, the values of $T$ did not differ much 
between Venus, Earth, and Mars for the cometary dust particles.
Below, $P_r=10^6 P$ and
$T_l$ is the longest lifetime of a particle in series of runs. 
For some runs at $\beta$$\ge$0.2,  all particles
starting close to perihelion  got hyperbolic orbits just after
starting from the comet. These runs are not included in the Table.

\begin{table}
\caption{Values of $T$, $T_l$ (in Kyr), and $P_r$  
(Venus=V, Earth=E, Mars=M)}

$ \begin{array}{lllccccc|lllccccc} 

\hline	

  & & & $V$ &  $E$ & $E$ & $M$ &  & & & & $V$ &  $E$ & $E$ & $M$ & \\


\beta & \Delta t_o && P_r & P_r & T &  P_r & T_l 
&\beta & \Delta t_o && P_r & P_r & T &  P_r & T_l \\ 
\hline

0.002 & 0    &S& 470 & 200 & 94.6 & 12 & 551 & 0.002 & 0    &L& 632 & 208 & 93.6 & 14 & 370 \\
0.002 & 0.25 &S& 408 & 156 & 84.3 & 13 & 20  &0.002 & 0.5  &S& 437 & 190 & 87.2 & 13 & 240 \\
0.004 & 0    &S& 370 & 148 & 62.9 & 8.9 & 213 & 0.004 & 0    &L& 303 & 139 & 66.0 & 9.0 &164 \\
0.004 & 0.25 &S& 430 & 160 & 55.0 & 9.3 & 109 & 0.004 & 0.5  &S& 235 & 140 & 56.3 & 8.1 & 108 \\
0.01  &  0  &S & 191 & 105 & 25.1 & 5.4 & 67 &0.01  &  0 &L & 386 & 163 & 28.5 & 6.4 & 80 \\ 
0.01  & 0.25&S & 238 & 86  & 24.2 & 4.2 & 59 &0.01  & 0.5&S  & 123 & 56  & 22.6 & 3.8 & 49 \\
0.05  & 0  &S  & 120 & 59  & 9.3  & 5.7 & 1070 &0.05  & 0  &L& 142 & 67  & 7.8  & 2.9 & 86 \\
0.05  & 0.25&S& 37 & 20 & 4.6 & 1.6 & 5 &0.05  & 0.5&S&   96 & 37   & 6.4 & 2.3 & 21 \\
0.1   & 0  &L& 23   &9.1   &6.1  & 0.6 & 229 &0.1   & 0.25 &S& 22 & 8.6  & 2.8  & 0.6 & 3 \\
0.1   & 0.5  &S& 13 & 6.6  & 2.7  & 0.47 & 3 & 0.2   & 0    &L& 7.4 & 3.5 & 3.6 &  0.27 & 119 \\
0.2   & 0.25 &S& 20 & 4.5 & 1.9 & 0.39 & 2 &0.2   & 0.5  &S& 12 & 3.2 & 1.6 & 0.22 & 2 \\
0.4   & 0.25 &S& 24 & 4.3  & 1.3  & 0.32 & 2 &0.4   & 0.5  &S& 13 & 3.5  & 1.4  & 0.22 & 2 \\

\hline
\end{array} $ 
\end{table}	

All particles with $\beta \le 0.01$ for $\Delta t_o = 0$ or
with $\beta \le 0.2$ for $\Delta t_o \ge 0.25$ collided with the Sun. 
Relatively large values of $T_l$ for 'S' runs
at $\beta=0.05$ and $\Delta t_o = 0$ 
and for 'L' runs at $\beta=0.1$ and $\Delta t_o = 0$ 
were due to the particles that reached 2000 AU from the Sun.
The values of $P_r$, $T$, and $T_l$ are greater for larger particles 
(i.e., for smaller $\beta$). The values of $P_r$ are greater for Venus 
than for Earth by a factor of 2 or more. 
Collision probabilities with Earth were greater by a 
factor of 10-20 than those with Mars and greater for particles
starting at perihelion than aphelion.
For the same values of $\beta$, the probability of cometary dust particles
colliding with a terrestrial planet was several times smaller 
than for asteroidal dust particles, mainly due to the 
greater eccentricities and inclinations of the cometary particles.
This difference is greater for larger particles.

The ratio of total times spent by cometary particles in inner-Earth, Aten, and Apollo 
orbits was about 1.5 : 1 : 2, but varied from run to run.

\section*{CONCLUSIONS}

We investigated collision probabilities of migrating asteroidal,
cometary, and kuiperoidal dust particles
with the terrestrial planets during the lifetimes of these particles.
These probabilities were considerably greater for larger 
asteroidal and cometary particles, 
which is in accordance with the analysis of microcraters.
The probability of collisions of cometary particles with the Earth
is smaller than for asteroidal particles, and this difference is greater
for larger particles.
 Almost all asteroidal particles with diameter $d$$\ge$4 $\mu$m
collided with the Sun. In almost all cases, inclinations $i$$<$50$^o$, and
at $a$$>$10 AU the maximum $i$ was less for larger asteroidal particles.
The spatial density of asteroidal particles decreases considerably at $a$$>$3 AU,
so the fraction of asteroidal particles among other particles beyond Jupiter's orbit is 
small. 
The peaks in the distribution of migrating asteroidal dust particles with semi-major axis
corresponding to the n:(n+1) resonances with Earth and Venus 
and the gaps associated with the 1:1 resonances with these planets
are more pronounced for larger particles.

\section*{ACKNOWLEDGEMENTS}

     This work was supported by NRC (0158730), NASA (NAG5-10776), INTAS (00-240), 
and RFBR (01-02-17540).

\end{document}